\documentclass[twocolumn]{article}
\usepackage[english]{babel}
\usepackage[utf8]{inputenc}
\usepackage{amsmath}
\usepackage{centernot}
\usepackage{amsfonts}
\usepackage{subcaption}
\usepackage{mathtools}
\usepackage{graphicx}
\usepackage{color}
\usepackage{authblk}
\usepackage{url}
\usepackage[switch]{lineno} 

\usepackage[colorlinks,citecolor=blue,linkcolor=blue,urlcolor=blue,bookmarks=false,hypertexnames=true]{hyperref}
\usepackage{geometry}
\graphicspath{{images/}}

\usepackage{listings}

\definecolor{dkgreen}{rgb}{0,0.6,0}
\definecolor{gray}{rgb}{0.5,0.5,0.5}
\definecolor{mauve}{rgb}{0.58,0,0.82}
\date{}

\title{Climate Change Modelling at Reduced Float Precision with Stochastic Rounding}
\author{Tom Kimpson\thanks{E-mail: tom.kimpson@physics.ox.ac.uk},
	E. Adam Paxton, 
	Matthew Chantry,
	and Tim Palmer}
\affil{\small \textit{University of Oxford, Oxford, UK}\normalsize}

\begin{document}

\maketitle

\begin{abstract}
Reduced precision floating point arithmetic is now routinely deployed in numerical weather forecasting over short timescales. However the applicability of these reduced precision techniques to longer timescale climate simulations - especially those which seek to describe a dynamical, changing climate - remains unclear. We investigate this question by deploying a global atmospheric, coarse resolution model known as SPEEDY to simulate a changing climate system subject to increased $\text{CO}_2$ concentrations, over a 100 year timescale. Whilst double precision is typically the operational standard for climate modelling, we find that reduced precision solutions (Float32, Float16) are sufficiently accurate. Rounding the finite precision floats stochastically, rather than using the more common ``round-to-nearest" technique, notably improves the performance of the reduced precision solutions. Over 100 years the mean bias error (MBE) in the global mean surface temperature (precipitation) relative to the double precision solution is $+2 \times 10^{-4}$K ($-8 \times 10^{-5}$ mm/6hr) at single precision and $-3.5\times 10^{-2}$ K($-1 \times 10^{-2}$ mm/6hr) at half precision, whilst the inclusion of stochastic rounding reduced the half precision error to +1.8 $\times 10^{-2}$ K ($-8 \times10^{-4}$ mm/6hr). By examining the resultant climatic distributions that arise after 100 years, the difference in the expected value of the global surface temperature, relative to the double precision solution is $\leq 5 \times 10^{-3}$ K and for precipitation $8 \times 10^{-4}$ mm/6h when numerically integrating at half precision with stochastic rounding. Areas of the model which notably improve due to the inclusion of stochastic over deterministic rounding are also explored and discussed. Whilst further research is necessary to extended these results to more complex and higher resolution models, they indicate that reduced precision techniques and stochastic rounding could be suitable for the next generation of climate models and motivates the use of low-precision hardware to this end.
\end{abstract}

\section{Introduction}
The use of reduced precision floating point arithmetic in place of the conventional 64 bit float (i.e IEEE-754 Float64 or `double') has gained recent attention in numerical weather modelling \cite{Chantry2019,Hatfield2019} for the potential to reduce computational demands on otherwise intensive processes. This saving in computational expense can then be reinvested in e.g. increasing model resolution, hence improving the accuracy of forecasts \cite{Lang2021}. For this reason, reduced precision methods are now used in operational weather forecasts; for example, the European Centre for Medium-Range Weather Forecasts runs the atmospheric component of the Integrated Forecast System (IFS) in single (32 bit float) precision \cite{ECMWF, Vana2016}. Weather forecasting centres in both the UK \cite{MetOffice} and Switzerland \cite{SWISS} have also explored the benefits of reduced precision arithmetic. Beyond numerical weather forecasting, reduced precision methods are now routinely deployed in neural network models \cite{Hopkins2020,Rehm2021,Gupta2021,Graphcore2022}, which motivates the development of dedicated CPU and accelerator hardware and architectures \cite{NVIDIA,TF} that can take advantage of reduced precision formats. These advances in hardware may in turn also be exploited by the numerical weather modelling community. \newline

\noindent Given the success of reduced precision methods in numerical weather prediction and the emergence of corresponding hardware, it is also of interest to investigate the use of the same techniques over longer timescales; from weather to climate modelling. This question was first investigated in \cite{Paxton2021} (`Pax21' hereafter) who explored the rounding error introduced as a result of using reduced precision for a global atmospheric climate model. They demonstrated that the rounding error using 12 significand bits was negligible and did not disrupt the long term climatology, whilst model accuracy could be maintained with acceptably minor errors emerging at 10 significand bits. They also briefly consider alternative rounding modes, beyond the standard  deterministic ``round-to-nearest". In particular the introduction of stochastic rounding appeared to have a modest effect to partially mitigate the rounding errors introduced at reduced precision. \newline

\noindent Motivated by these results, in this work we build on the previous study of Pax21 to investigate the efficacy of reduced precision when applied to modelling a changing climate system, subject to increased CO$_2$ levels. The global atmospheric modelling of Pax21 used a constant-in-time sea surface temperature (SST) anomaly field, such that the boundary conditions were annually periodic and the system admitted a well defined, invariant climate distribution. Whilst this enabled them to obtain more stringent bounds on the rounding error through minimizing any external variability due to e.g. non-stationarity, for the purposes of applying reduced precision arithmetic to accurate climate models it is important to consider a changing climate, particularly in response to variations in atmospheric CO$_2$. \newline

\noindent In this work we will deploy the same global atmospheric model as used in Pax21, but now remove these previous constraints and instead use a non-stationary SST anomaly field along with an instantaneous quadrupling of the atmospheric CO$_2$. This enables us to effectively perform a climate change experiment, where we will focus on two key questions:
\begin{enumerate}
	\item Can we reproduce the climate change signal at reduced floating point precision?
	\item Can stochastic rounding help mitigate any errors incurred at reduced float precision?
\end{enumerate}
The atmospheric modelling used here is an inherently simplified version of a full state-of-the-art climate model. However, definitive answers to the above questions using this simplified model would give strong indications as to the potential of reduced precision arithmetic and alternative float rounding methods when applied to more complex climate simulations, and may suggest fruitful avenues for increasingly accurate and computationally demanding climate modelling. \newline 

\noindent The paper is organised as follows. In Section \ref{sec2_floats} we give a short overview of round-to-nearest and stochastic rounding techniques for finite precision arithmetic. In Section \ref{sec3_climatechange} we describe the atmospheric model used and our climate change experiment in more detail, before running the experiment at different float precisions. We explore both the transient behaviour as the climate is changing and the ultimate invariant distribution (i.e. climate) that the solutions equilibrate to. A focused analysis on the impact of stochastic rounding and the potential of mixed precision solutions is presented in \ref{sec4_SRMP}, a general discussion on the experiment results and potential implications in Section \ref{sec5_discus}, before concluding remarks and suggestions for future directions of work.

\section{Rounding Floats} \label{sec2_floats}

In general, the result of arithmetic operations on a computer using finite precision arithmetic do not have an exact representation in that number system. Instead, it is necessary to map - or round - from the result to a representable format. The difference between the true operation result and the representable value is the well known rounding error of floating point arithmetic. The IEEE 754 standard \cite{IEEEstandard} uses a``round-to-nearest" (RN) mapping where the operation result is rounded to the nearest representable value.  It is fundamentally a deterministic process; a given result will always be mapped to the same representation. The error introduced by a RN rounding can be shown to be bounded as \cite{Higham2002, Connolly2021}
\begin{eqnarray}
	\mathcal{R}(x) = x(1+\delta), \, \, \delta \leq \epsilon/2
\end{eqnarray}
for machine epsilon $\epsilon$. \newline 

\noindent Alternatively, rather than deterministic RN, it is also possible to use stochastic rounding (SR). In this case the result of an arithmetic operation is mapped to a representable form as,
\begin{equation}
	\mathcal{R}(x) =
	\begin{cases}
		 \left \lceil{x}\right \rceil & \text{with probability $p(x)$}\\
		 \left \lfloor{x}\right \rfloor  & \text{with probability $1-p(x)$}\\
	\end{cases}       
\end{equation}
where 
\begin{equation}
	p(x) = \frac{x-\left \lfloor{x}\right \rfloor}{\left \lceil{x}\right \rceil -\left \lfloor{x}\right \rfloor}
\end{equation}
\cite{SR1,SR2}. As noted in \cite{Connolly2021}, there are some standard properties that hold under RN which no longer hold under stochastic rounding. For instance, under SR, generally $\mathcal{R}(-x) \neq -\mathcal{R}(x)$  (lack of rounding symmetry due to stochasticity) and $x \leq y \centernot\implies \mathcal{R}(x) \leq \mathcal{R}(y)$ (non-monotonicity of rounding). Moreover, the rounding error is now bounded as, 
\begin{eqnarray}
	\mathcal{R}(x) = x(1+\delta),  \, \, \delta \leq \epsilon
\end{eqnarray}
a less stringent bound than for RN. Furthermore, since stochastic rounding naturally requires random number generation, it is likely more expensive than a deterministic process. In spite of these apparent drawbacks, SR has some important properties which make it advantageous from the perspective of both training neural networks e.g.\cite{Gupta2015,Na2017,Wang2018,Xia2021} and numerical solutions to differential equations e.g.  \cite{Hopkins2020,Fasi2021,Croci2022}.  In particular, SR is a statistically unbiased rounding method in that $\mathbb{E} \left(\mathcal{R}(x)\right) = x$. This also helps prevent numerical stagnation whereby when updating a number $x$ by some small quantity $\delta$ using round-to-nearest, $\mathbb{E} \left(\mathcal{R}(x+\delta)\right) = x$. Consequently the information carried in $\delta$ is lost, whereas it is preserved under SR. To take a relevant physical example, if during numerical climate modelling a small temperature tendency is always set to zero when using deterministic rounding this will affect the melting of e.g. permafrost. Whilst stochastic rounding has not traditionally been supported in hardware, the emerging trend of low precision arithmetic in machine learning has strongly motivated the development of new hardwares. For example, the new Graphcore Intelligence Processing Unit (IPU) - developed explicitly for machine learning applications - has inbuilt stochastic rounding support \cite{Jia2019} \cite{graphcore}. A comprehensive review of available and in development SR hardware is presented in \cite{CrociSRReview}.

 \section{Climate Change Experiment} \label{sec3_climatechange}
 
 \subsection{Overview}
 Throughout this work, following Pax21, we will use a global atmospheric model known as SPEEDY (Simplified Parameterizations PrimitivE Equation DYnamics, ver 41) \cite{SPEEDYOG} that has been modified to run in low precision \cite{SPEEDY}  utilizing a reduced precision emulator rpe v5 \cite{rpeV5}. SPEEDY is an intermediate complexity, coarse resolution model that uses a spectral transform dynamical core with a spectral resolution of T30,  a 96x48 Gaussian grid, a 40 minute timestep and eight vertical sigma levels.  The numerical integration timestep itself uses a leapfrog scheme in conjunction with a Robert–Asselin–Williams filter \cite{RAW}. Daily updated boundary conditions are determined by taking monthly averaged values (e.g. for sea ice) and then linearly interpolating. We use the same set of initial conditions as described in Pax21. \newline 
  
\noindent We are fundamentally interested in being able to model a changing climate. A widely used metric in climate modelling is the equilibrium climate sensitivity (ECS) which describes the change in the global mean surface temperature after an instantaneous doubling of CO$_2$ compared to pre-industrial levels. In order to evaluate the sensitivity of a climate model to CO$_2$ forcings and determine the ECS, it is common practice to use a ``4xCO2" experiment where a spun-up pre-industrial run is subject to an abrupt quadrupling of the atmospheric CO$_2$ concentration; the initial Diagnostic, Evaluation and Characterization of Klima (DECK) requirements for participation in Phase 6 of the CMIP includes a 4xCO2 experiment \cite{Eyring2016}, whilst a previous EC-Earth experiment has explored the change in the ECS when progressing from CMIP5 to CMIP6 models  \cite{ECEarth}. For our climate change experiment we follow the same example and instantaneously quadruple the CO$_2$ concentration. Specifically, we set the absorptivity of air in the CO$_2$ band to be 21, as a rough estimate of a value 4 times the pre-industrial level. This quadrupling is advantageous for our purposes since it subjects the reduced precision arithmetic to a more strenuous test over simply doubling the CO$_2$ concentration. Rather than holding the SST field constant as in Pax21, we take the SST from the corresponding EC-Earth experiment, re-gridded to match the SPEEDY grid.  As a shorthand we will refer to this system with non-constant SST and increased CO$_2$ concentrations  as a ``4xCO2" world. \newline

\noindent We run the experiment at the following precisions, where we truncate the significand bits only:
\begin{itemize}
	\item Double (Float64) - 53 significand bits, 11 exponent bits 
	\item `Single' (Float32) - 23 significand bits, 11 exponent bits
	\item `Half' - 10 significand bits, 11 exponent bits
\end{itemize}
For the solutions at half precision we will consider both round-to-nearest and stochastic rounding methods. We will use the notation $X\_Y$ to refer to a solution with $X$ significand bits using a rounding method $Y$, e.g. $10\_\text{SR}$ labels the half precision solution, using stochastic rounding.

\subsection{Transient effects of climate change}

\begin{figure} 
	\includegraphics[width=\columnwidth]{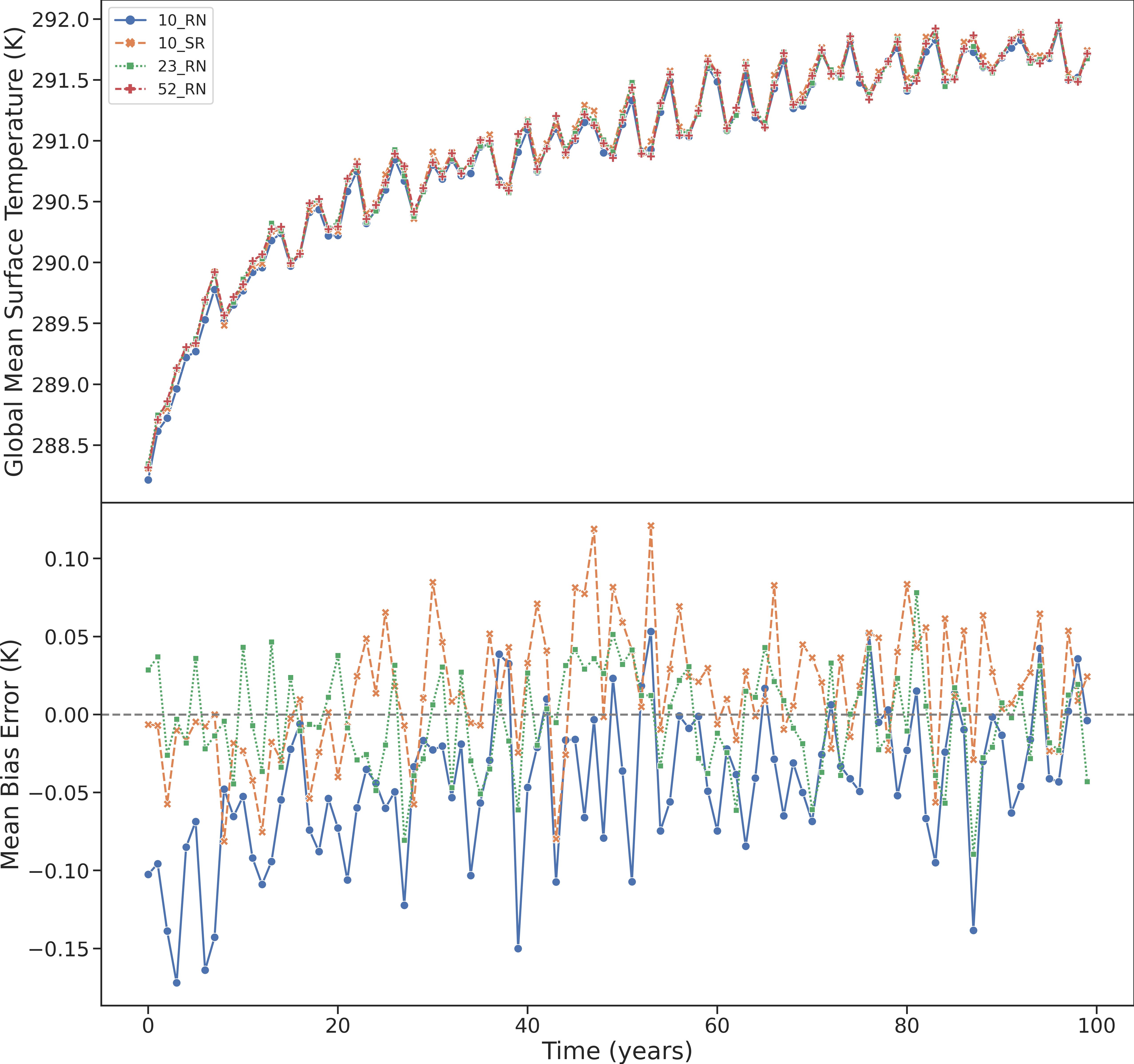}
	\caption{Time evolution of the global mean surface temperature over 100 years for a 4xCO$_2$ world (top panel), and the associated mean bias error relative to the double precision solution (bottom panel). Each data point represents an average over a 1 year timescale. The reduced precision 10\_RN solution performs well, but is generally offset from the ground truth, especially at early times, with a mean bias error averaged over the entire integration period of $-0.044$ K. The addition of stochastic rounding (10\_SR) notably improves the solution, with mean bias error of $0.014$ 	K.}
	\label{fig:single_timeseries}
\end{figure}

Initially, consider as a demonstrative example a particular SPEEDY integration over 100 years from an single initial condition, for the 4xCO2 world. The evolution of the global mean surface temperature (i.e. the latitude-weighted global average of the surface temperature at every grid point) over a 100 year timescale is presented in Fig. \ref{fig:single_timeseries}, along with the error relative to the full double precision solution, which we take as our ground truth. 
\begin{figure} 
	\includegraphics[width=\columnwidth]{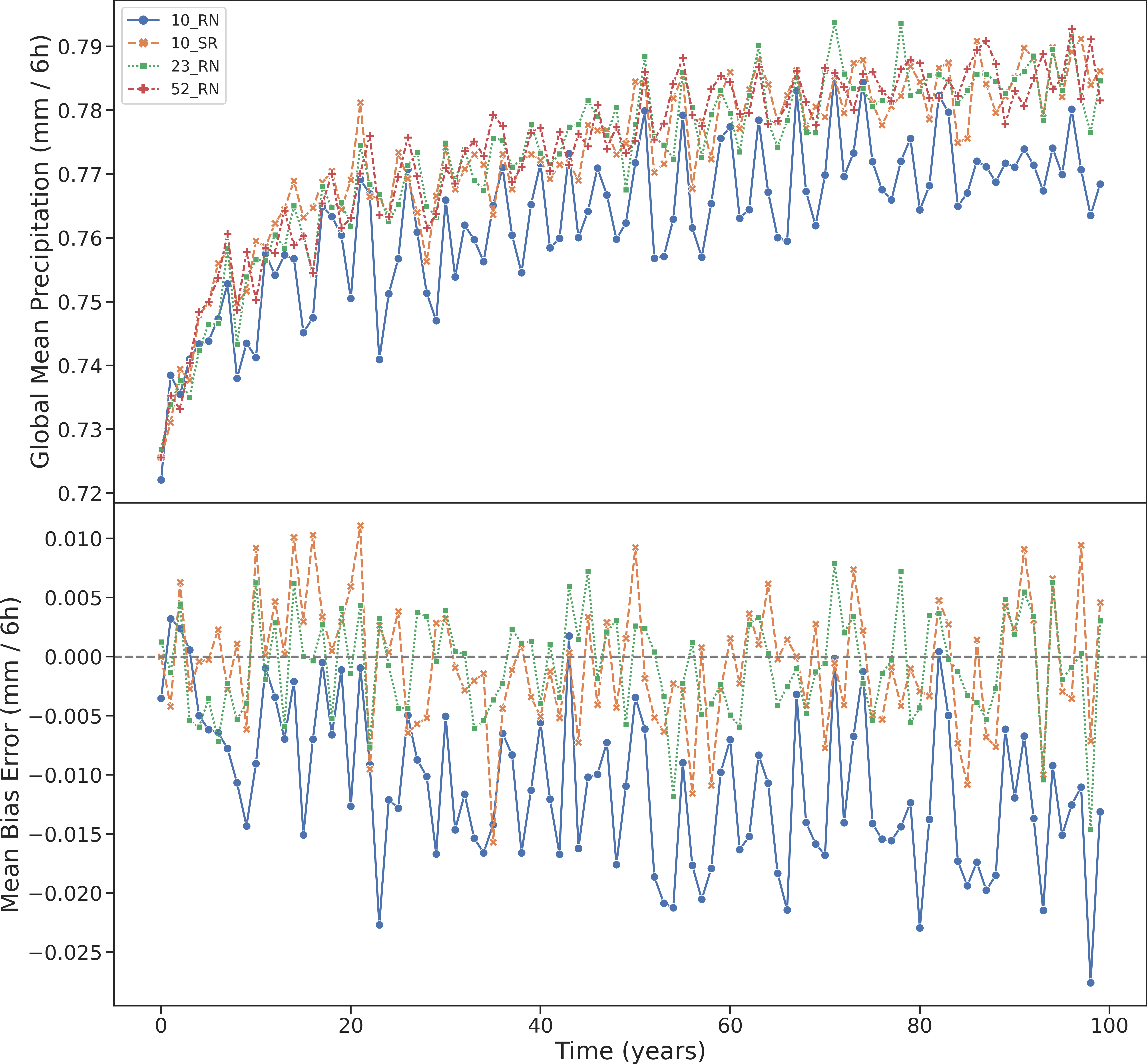}
	\caption{As Fig. \ref{fig:single_timeseries}, for global mean total precipitation (sum of large-scale and convective precipitation). The 23\_RN and 10\_SR solutions are generally faithful to the full double precision solution, with $|\text{MBE}| \le 6 \times 10^{-4}$ mm/6h. Conversely the 10\_RN  solution can be seen to diverge from the truth over time.}
	\label{fig:single_timeseries_precip}
\end{figure}
\noindent We can see that for a 4xCO$_2$ world the global mean surface temperature generally increases logarithmically by just under $\sim 4$K over 100 years. Against the background of this general increase, there is also a shorter timescale, $\sim \mathcal{O} (\rm years)$, oscillation in the temperature. Both the overall profile and the magnitude of the true (Float64) solution are generally well-reproduced by all the reduced precision solutions. Whilst there is clearly a high degree of volatility year-on-year in the error, the relative accuracy in the reduced precision solutions can be clearly discerned. Single precision exhibits the best performance, tracking the true solution accurately and without bias: averaging over the 100 year period, the mean bias error (MBE) is $- 3.6 \times 10^{-3}$ K and mean absolute error (MAE) $2.7 \times 10^{-2}$ K. The half precision solution using round-to nearest is less performant, with an MAE of $5.2 \times 10^{-2}$ K , i.e. approximately twice as bad as 23\_RN. Additionally the solution is much more strongly biased than the single precision solution, with MBE $= - 4.4 \times 10^{-2}$ K. This is evident from Fig. \ref{fig:single_timeseries} where the 10$\_$RN solution is distinctly offset from the true time series, consistently underestimating the surface temperature whilst still following the general profile. However, the addition of stochastic rounding to the half precision solution notably improves the performance, with MBE $ = +1.4 \times 10^{-2}$ K and MAE $= 3.4 \times 10^{-2}$. Similar results are observed if we instead consider global mean precipitation (Fig. \ref{fig:single_timeseries_precip}); the double precision solution is well-reproduced at both single precision (MBE $ \sim - 8 \times 10^{-4}$ mm/6h) and half precision with stochastic rounding (MBE $\sim - 7 \times 10^{-4}$ mm/6h) whilst the half precision solution using round to nearest is distinctly less accurate (MBE $ \sim -1.2 \times 10^{-2}$ mm/6h). \newline  

\noindent From these single runs a clear hierarchy of solutions emerges, with 23\_RN $>$ 10\_SR $>$ 10\_RN. Whilst the 10\_SR solution is not quite as good as the single precision solution it still exhibits a generally improved performance over the 10\_RN solution which has an identical number of information bits. Moreover, the 10\_SR solution is broadly comparable with the 23\_RN solution, despite carrying 13 fewer bits of significand information. Furthermore, the saved computer time by operating at a reduced precision could be reinvested to e.g. increase the vertical resolution which could have a stronger effect on improving the accuracy than the small degradation of 10\_SR. This is an encouraging initial result c.f. the two questions we posed at the start of this paper; the climate change signal can be accurately reproduced at low precision and stochastic rounding generally  improves  the solution - at least for this demonstrative single run using a simplified atmospheric model. \newline 

\begin{figure*}[h!]
	\includegraphics[width=0.98\textwidth]{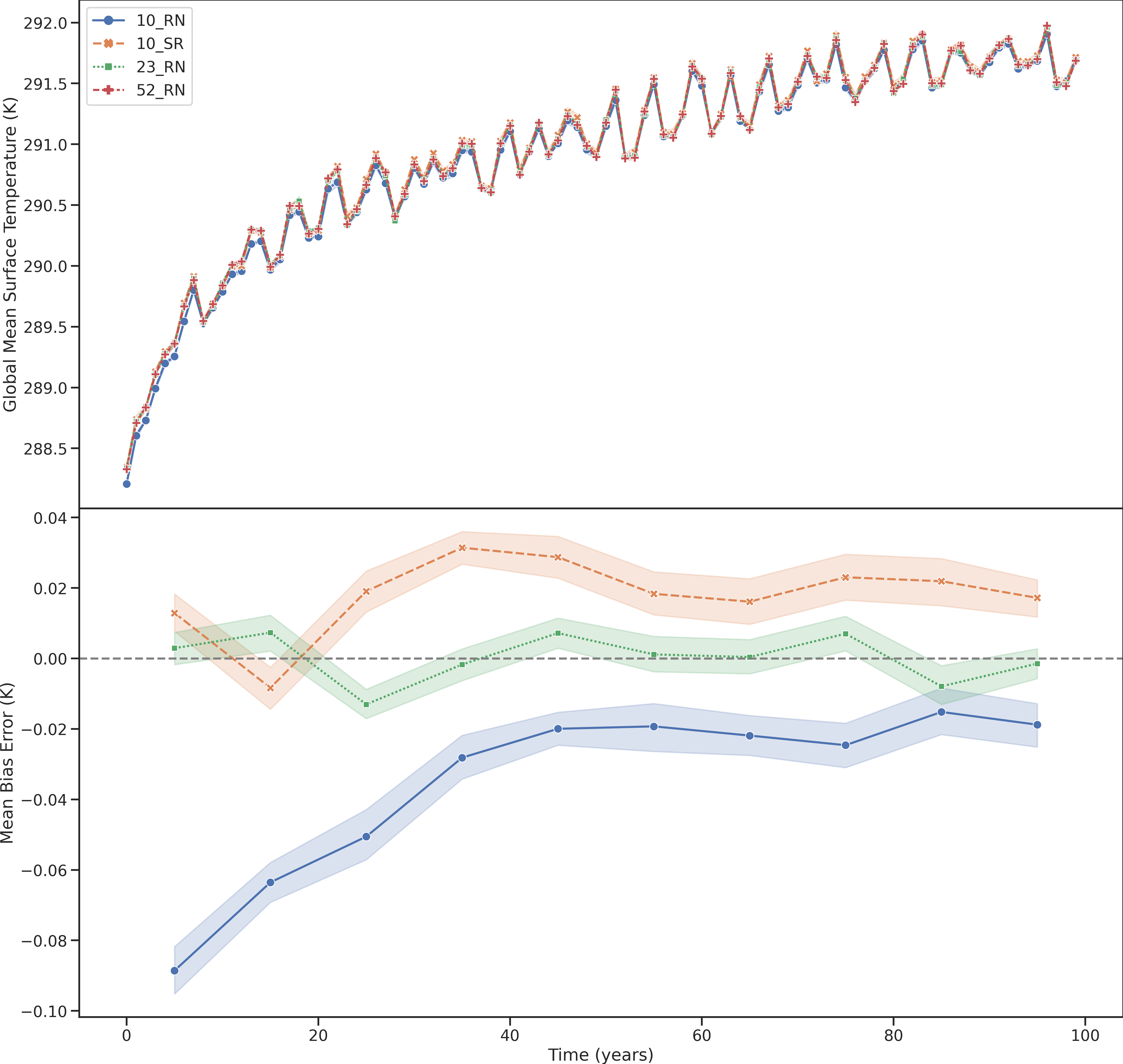}
	\caption{Time evolution of the global mean surface temperature over a 100 years for a 4xCO$_2$ world, averaged over 5 ensemble members (top panel) and the ensemble bias error (bottom panel). The ensemble temperature dispersion is present, but too small to be seen on this scale. The ensemble bias is presented as a decadal average so as to smooth out the volatility and display the general trends.}
	\label{fig:ensemble_timeseries}
\end{figure*}

\begin{figure*}[h!]
	\includegraphics[width=0.98\textwidth]{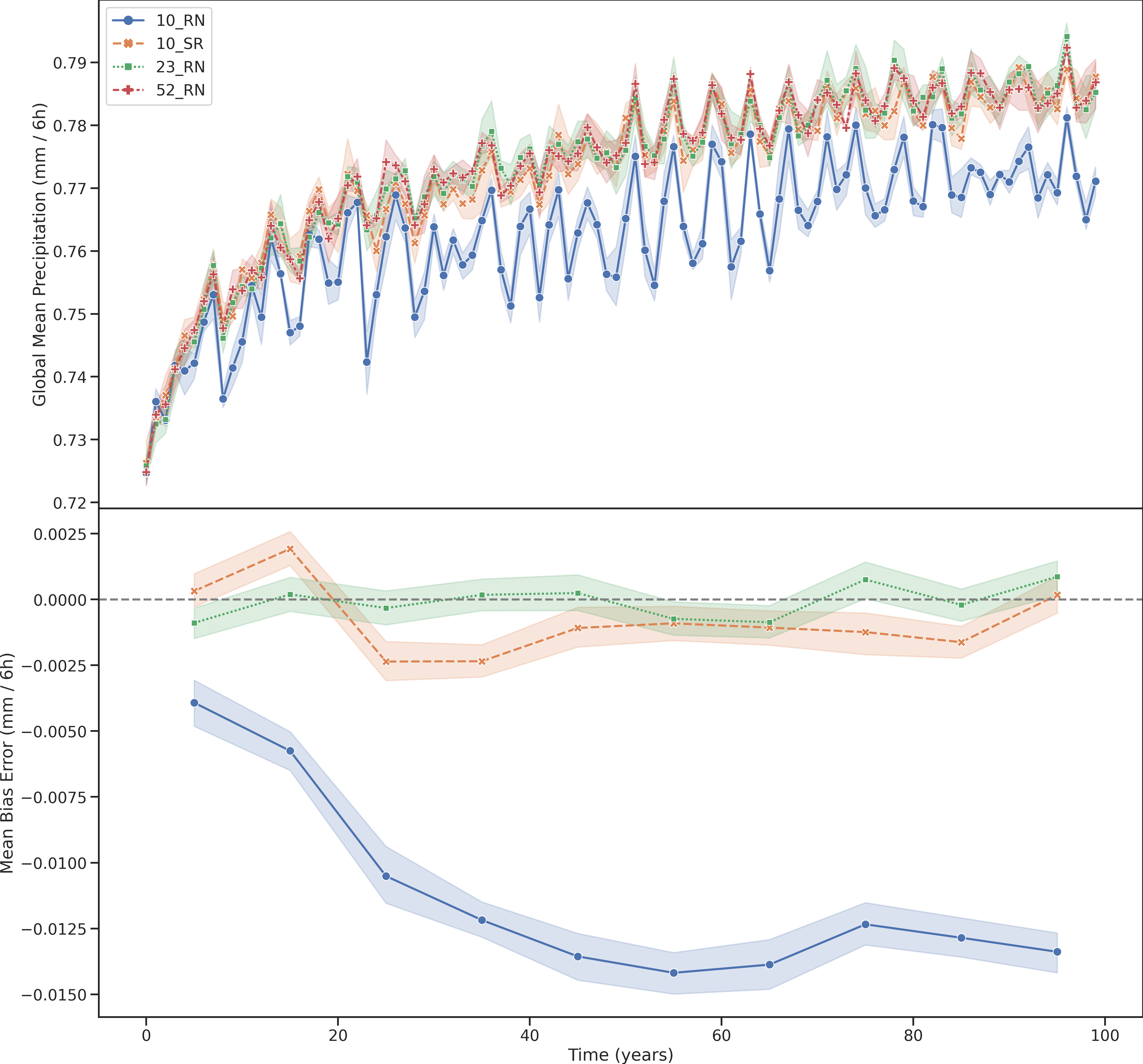}
	\caption{As Fig. \ref{fig:ensemble_timeseries}, for global mean total precipitation. The 23$\_$RN and 10$\_$SR solutions accurately track the ground truth to within 0.0025 mm/6h over the entire 100 years, whilst the 10$\_$RN gradually diverges over time.}
	\label{fig:ensemble_timeseries_precip}
\end{figure*}

\noindent Building upon this single run, we can now consider an ensemble of initial conditions and explore the transient behaviour across all ensemble members. We use 5 members per ensemble for each precision. The results are presented in Figs. \ref{fig:ensemble_timeseries}, \ref{fig:ensemble_timeseries_precip} for temperature and precipitation respectively. The behaviour is generally similar to our previous integration from a single initial condition; all the reduced precision solutions generally perform well, faithfully reproducing the true double precision solution to within $\sim$ 0.1 K for temperature and $\sim 0.015$ mm/6h for precipitation. The same hierarchy of solutions is also evident, with the half precision stochastic rounding solution outperforming round-to-nearest; for temperature (precipitation) at 10\_SR the MBE is $1.8 \times 10^{-2}$ K ($ -8.2 \times 10^{-4}$ mm/6h) whilst for 10$\_$RN it is $-3.5 \times 10^{-2}$ K ($ -1.1 \times 10^{-2}$ mm/6h). That is to say, for temperature stochastic rounding results in a solution that is approximately twice as accurate as the round to nearest, whilst for precipitation it is ${\sim}14$x more accurate. It is also notable that in both cases the 10$\_$RN profile generally underestimates the true solution, characteristic of numerical stagnation. For temperature, this underestimation is especially prominent at early times when the global surface temperature is increasing rapidly due to the changing climate. The generally logarithmic nature of the true temperature profile as the rate of warming decreases as the climate starts to equilibrate dampens the impact of this stagnation, giving the 10$\_$RN  solution a chance to ``catch up" at later times.  This then suggests that using as an accuracy metric the 100 year average of the error is perhaps too generous  to the 10$\_$RN solution which does not accurately  following the true solution over all time in the way that 23$\_$RN and 10$\_$SR do, but instead has its score buoyed at later times. For instance if we calculate the MBE over the initial 25 years rather than the full 100 year integration, then the MBE for 10$\_$RN is $7.2 \times 10^{-2}$ K, for 10$\_$SR is it $4.9 \times 10^{-3}$ K.  Conversely the precipitation is not subject to such rapid increases and the accuracy of the 10$\_$RN solution gradually decays over time. Whilst the 10$\_$SR solution performs well, and can be seen to be superior to 10$\_$RN, it is worth acknowledging that it is generally slightly less performant than 23$\_$RN which has MBE over 100 years of $2 \times 10^{-4}$ K and $8 \times 10^{-5}$ mm/6h for temperature and precipitation respectively. The 10$\_$SR and 23$\_$RN solutions are highly comparable for precipitation, with 23$\_$RN a touch more accurate, whilst the superiority of 23$\_$RN over 10$\_$SR is much more evident for temperature. However, 10$\_$SR achieves almost on-par performance over 100 years with 13 fewer bits of significand information, and indeed 43 fewer significand bits than the double precision solution! More broadly the strong performance of single precision suggests that similar to numerical weather modelling, climate models could potentially be straightforwardly run at Float32. \newline

\begin{figure}
	\subfloat[\label{fig:molmap1}]{\includegraphics[clip,width=\columnwidth]{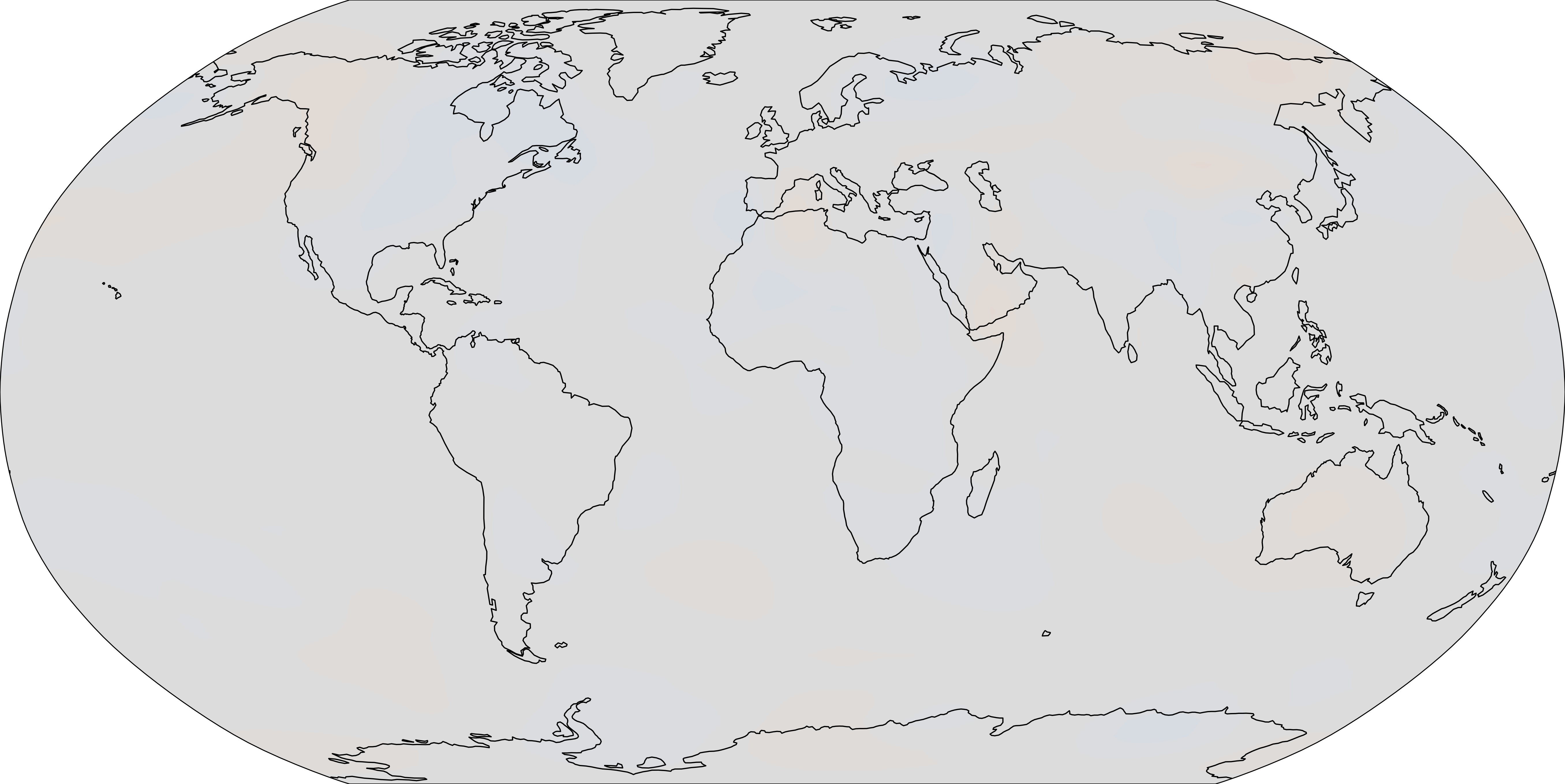}}
	
	\subfloat[\label{fig:molmap2}]{\includegraphics[clip,width=\columnwidth]{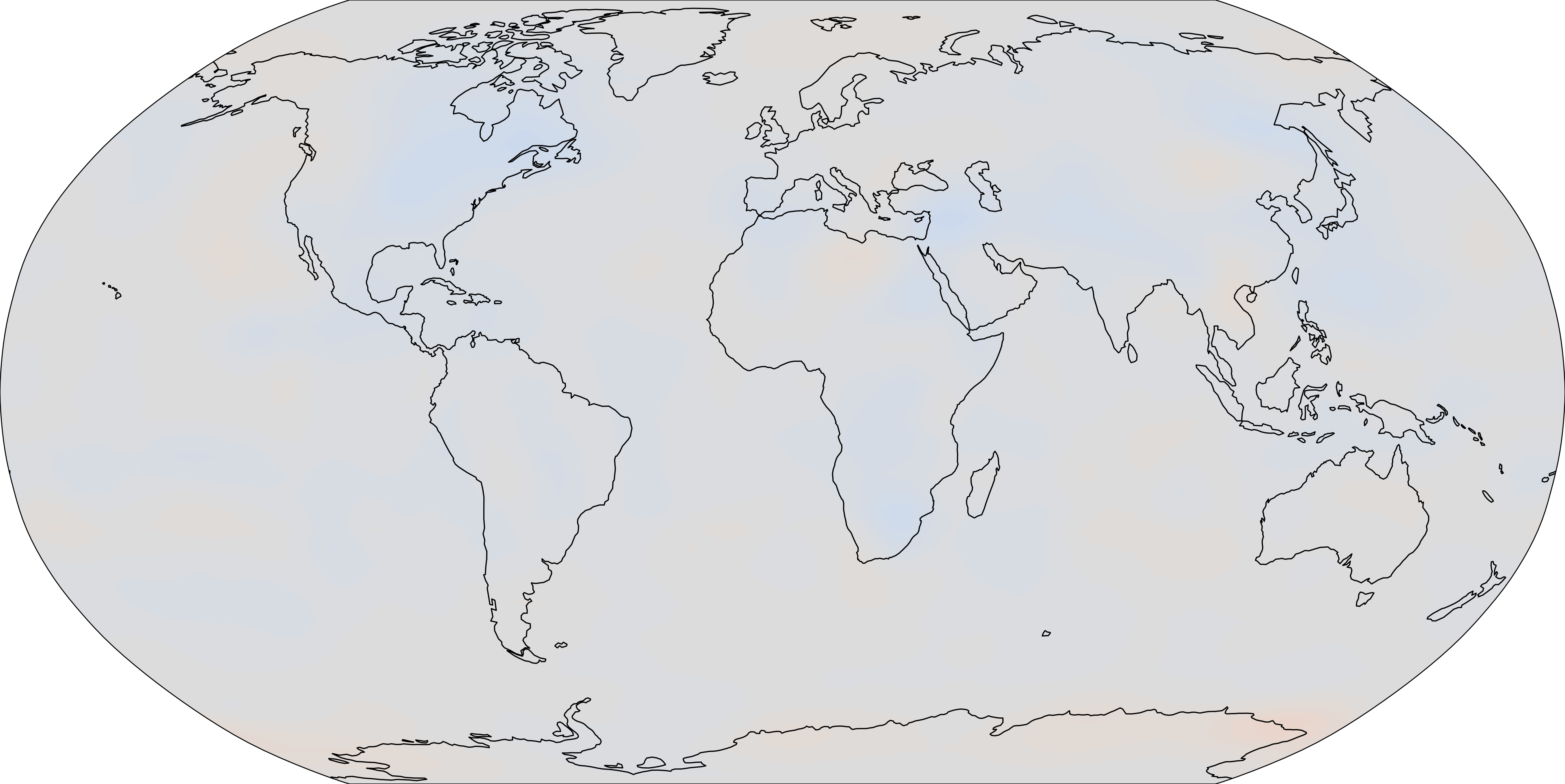}}
	
	\subfloat[\label{fig:molmap3}]{\includegraphics[clip,width=\columnwidth]{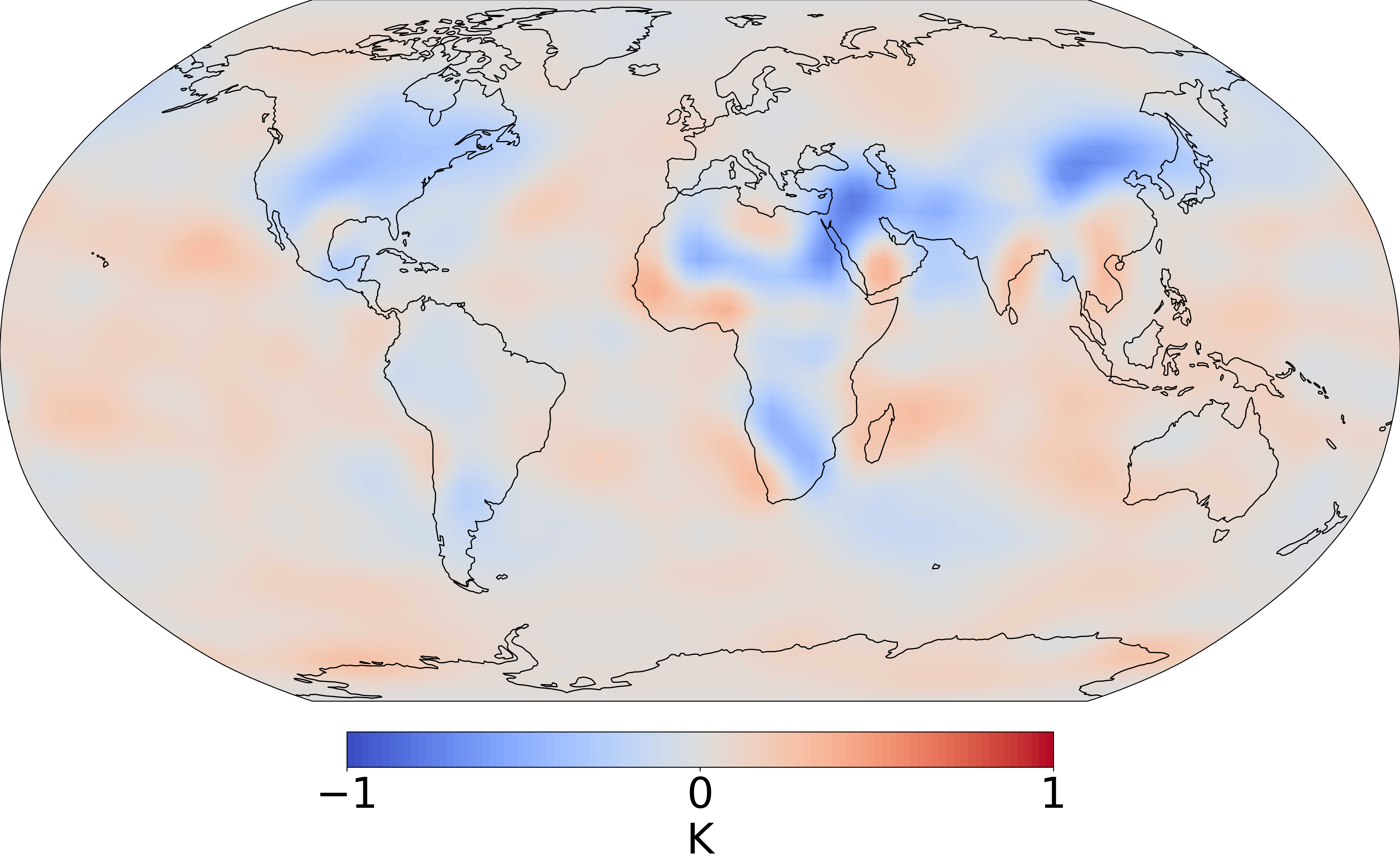}}
	
	\caption{Global distribution in the difference in temperature between the control ensemble and the competitor ensemble, averaged over 100 years at \textit{(a)} double precision, \textit{(b)} half precision with stochastic rounding, \textit{(c)} half precision with round-to-nearest. Analogous to the temperature time series, we can see that 10$\_$SR is globally an improvement over 10$\_$RN and indeed is highly similar to the 52$\_$RN solution.	 } 
	\label{fig:molmap}
\end{figure}

\begin{figure}
	\subfloat[\label{fig:molmap1_precip}]{\includegraphics[clip,width=\columnwidth]{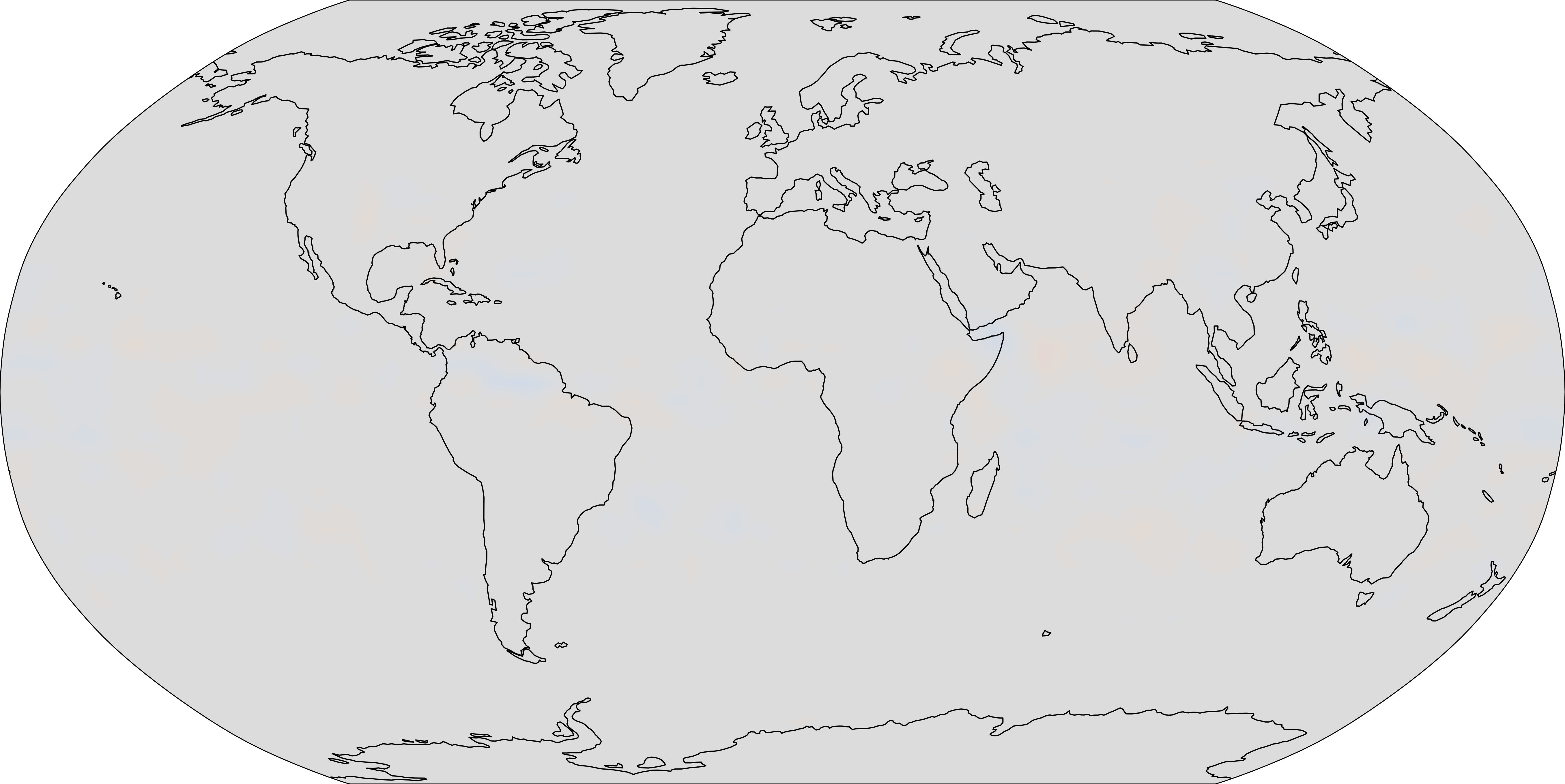}}
	
	\subfloat[\label{fig:molmap2_precip}]{\includegraphics[clip,width=\columnwidth]{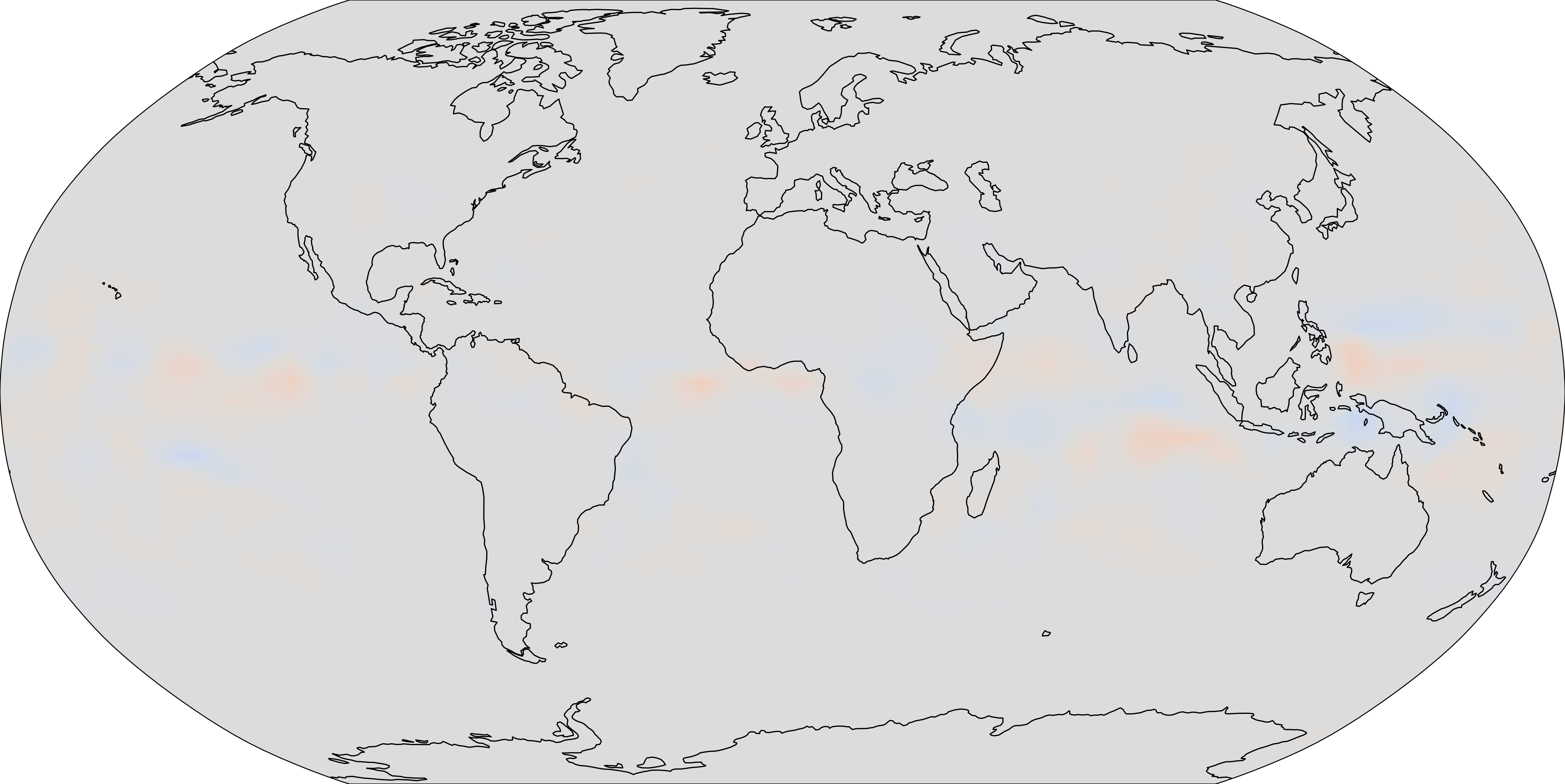}}
	
	\subfloat[\label{fig:molmap3_precip}]{\includegraphics[clip,width=\columnwidth]{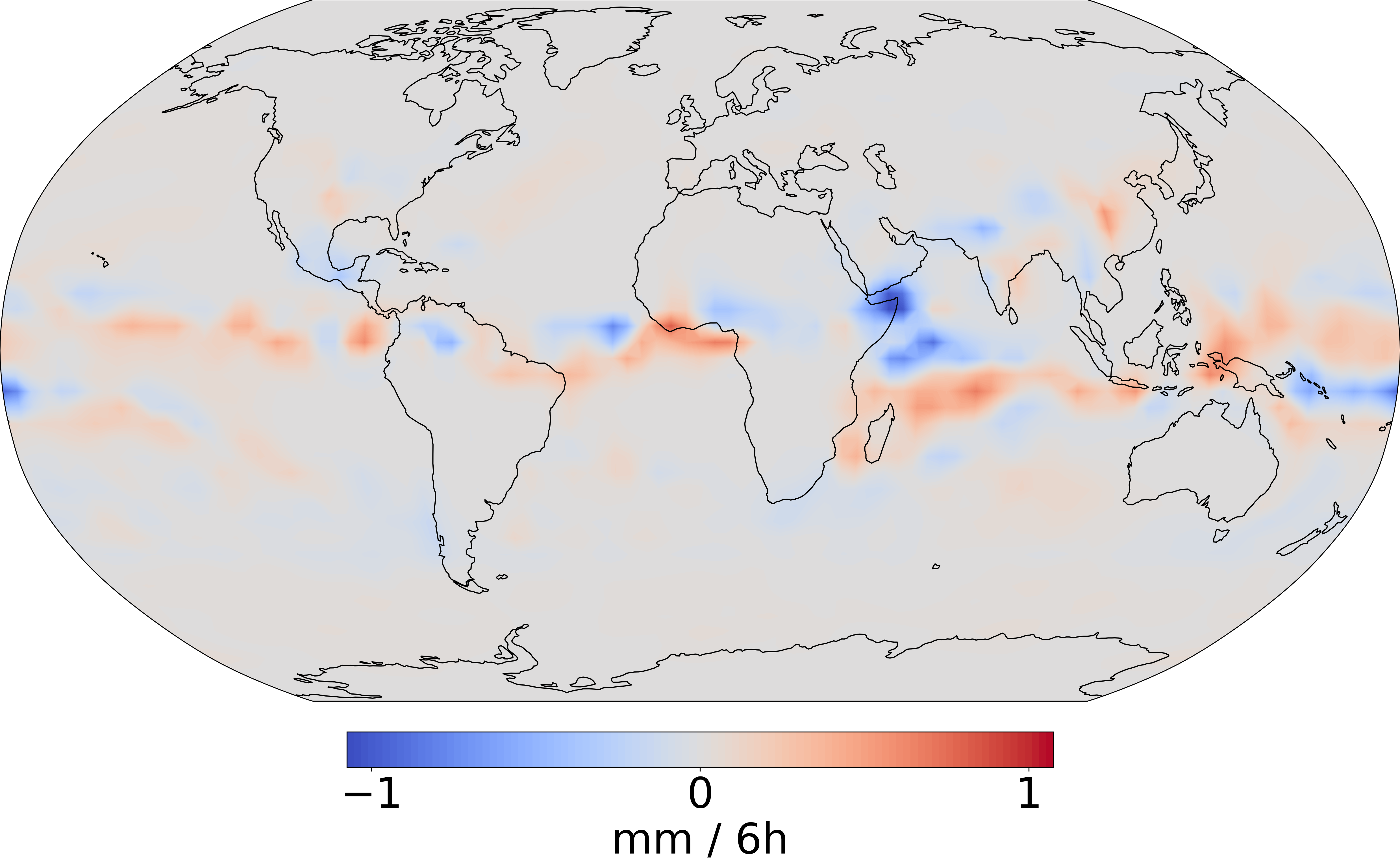}}
	
	\caption{As Fig. \ref{fig:molmap}, but for precipitation (sum of large scale precipitation and convective precipitation). The largest errors for the two reduced precision solutions are seen in the tropics where the precipitation is naturally greater. Again the 10$\_$SR solution outperforms the 10$\_$RN solution. \vspace{9mm} } 
	\label{fig:molmap_precip}
\end{figure}

\noindent In addition to considering the global average time series, we can also explicitly investigate the spatial behaviour of the reduced precision solutions; are there particular geographic areas where 10\_RN is failing but 10\_SR succeeds? In order to do this we compare our `competitor' ensembles at each precision against a separate control ensemble, which is composed of 5 members at double precision. Figs. \ref {fig:molmap}, \ref {fig:molmap_precip} show the spatial distribution in the difference between the competitor and control  ensembles, averaged over 100 years, for temperature and precipitation respectively. In both cases the  10$\_$SR outperforms  10$\_$RN globally. For the surface temperature, the 52$\_$RN solution is relatively uniform on this $\sim $ K scale, showing that the inherent chaotic variability in the mean surface temperatures is sub-Kelvin. The 10$\_$SR map is also relatively featureless, and is highly comparable to the double precision solution. Conversely, the 10$\_$RN solution shows large overestimates relative to the control group mainly over land, primarily in North Africa and the Middle east as well as North America and South Africa, and large underestimates over the sea. For precipitation, the same trend regarding the relative performance of 10$\_$SR and 10$\_$RN is true, but now the errors are localised to the regions with the largest rates of precipitation, namely the tropics. In this case neither solution exhibits a particularly strong over/under estimate bias in particular regions. 

\subsection{Invariant Climate Distribution}

The preceding analysis was concerned with the transient effects of climate change. During this time the climates are ill-defined and it is not possible to consistently define some invariant distribution that can then be compared at different float precisions - the approach used in Pax21 for constant SSTs and no external forcings. However, it is of interest to be able to examine the invariant distribution that arises over time. i.e. the climate which is settled upon after climate change has finished. To this end we now take the system after 100 years of climate change and re-impose constant SST fields, analogous to Pax21, and integrate for a further 10 years.  We will construct a control ensemble $e^{\text{control}}$ at double precision and a competitor ensemble $e^{\text{competitor}}$ at each of the double/single/half float precisions. We use the same set of initial conditions $i_j$ for $j=1,...,10$ as described in Pax21. These initial conditions were obtained via integrating a system from rest for 11 years at double precision with annually periodic boundary conditions, discarding the first year, and taking the start of each subsequent year as the $j^{\text{th}}$ initial condition. The control ensemble is constructed from the set of initial conditions $i_1...i_5$ whilst a competitor ensemble  is constructed from $i_6, ..., i_{10}$. We will use the notation $e_j$ to refer to the $j^{th}$ member of an ensemble, which has initial condition $i_j$.  \newline 

\noindent In order to compare the control and competitor ensemble probability distributions we adopt as our key metric the Wasserstein distance (WD). The WD intuitively defines a distance between two distributions as the optimal cost, with respect to some cost function $c(x,y)$,  of transporting a probability mass from position $x$ to position $y$. The $p$-Wasserstein distance between two distributions $\mu, \nu$ is
\begin{eqnarray}
	W_p(\mu,\nu)= \left( \inf_{\gamma \in \Gamma(\mu, \nu)}  \int c(x,y)^p d \gamma (x,y)\right)^{1/p}
\end{eqnarray}
where $\gamma(x,y)$ is the transport plan, $\Gamma(\mu, \nu)$ is the set of all couplings of $\mu$ and $\nu$ and $c(x,y)$ is the cost function. It is well-known \cite{Dudley} that the Wasserstein distance suffers from the curse of dimensionality for dimensions $d \geq 3$:
\begin{eqnarray}
	\mathbf{E} [W_1(\mu_n,\mu)] \asymp n^{-\frac{1}{d}}
\end{eqnarray}
To sidestep this issue, we marginalize onto the distributions spanned by the individual gridpoints and take the WD between these 1-dimensional distributions. Such an approach is also adopted in \cite{Paxton2021,Vissio2020}. We also take $p=1$ and $c(x,y) = |x-y|$ such that the WD values quoted in this work are given by
\begin{eqnarray}
	W_1(\mu,\nu)=  \inf_{\gamma \in \Gamma(\mu, \nu)} \int |x-y| d \gamma (x,y)
\end{eqnarray}
From the perspective of climate modelling, the WD has a nice physical interpretation given by the Monge-Kantorovich duality:
\begin{eqnarray}
	| \mathbf{E}(X_{\mu} )-\mathbf{E}(Y_{\nu} ) | \leq W_1(\mu, \nu) \label{eq:WDdefn}
\end{eqnarray}
i.e. the difference in the expected value of some random variable between two distributions is no bigger than the WD value itself. More explicitly, if the WD for surface temperature between two distributions is 1K, then this tells us that the difference in the expected temperature between the distributions is no greater than 1K. For a full review of the Wasserstein distance, including favourable properties and comparison with other metrics we refer the reader to the Appendix of Pax21. \newline 
 \begin{figure*}
	\subfloat[\label{fig:wasserstein_temperature}]{\includegraphics[width=0.48\textwidth]{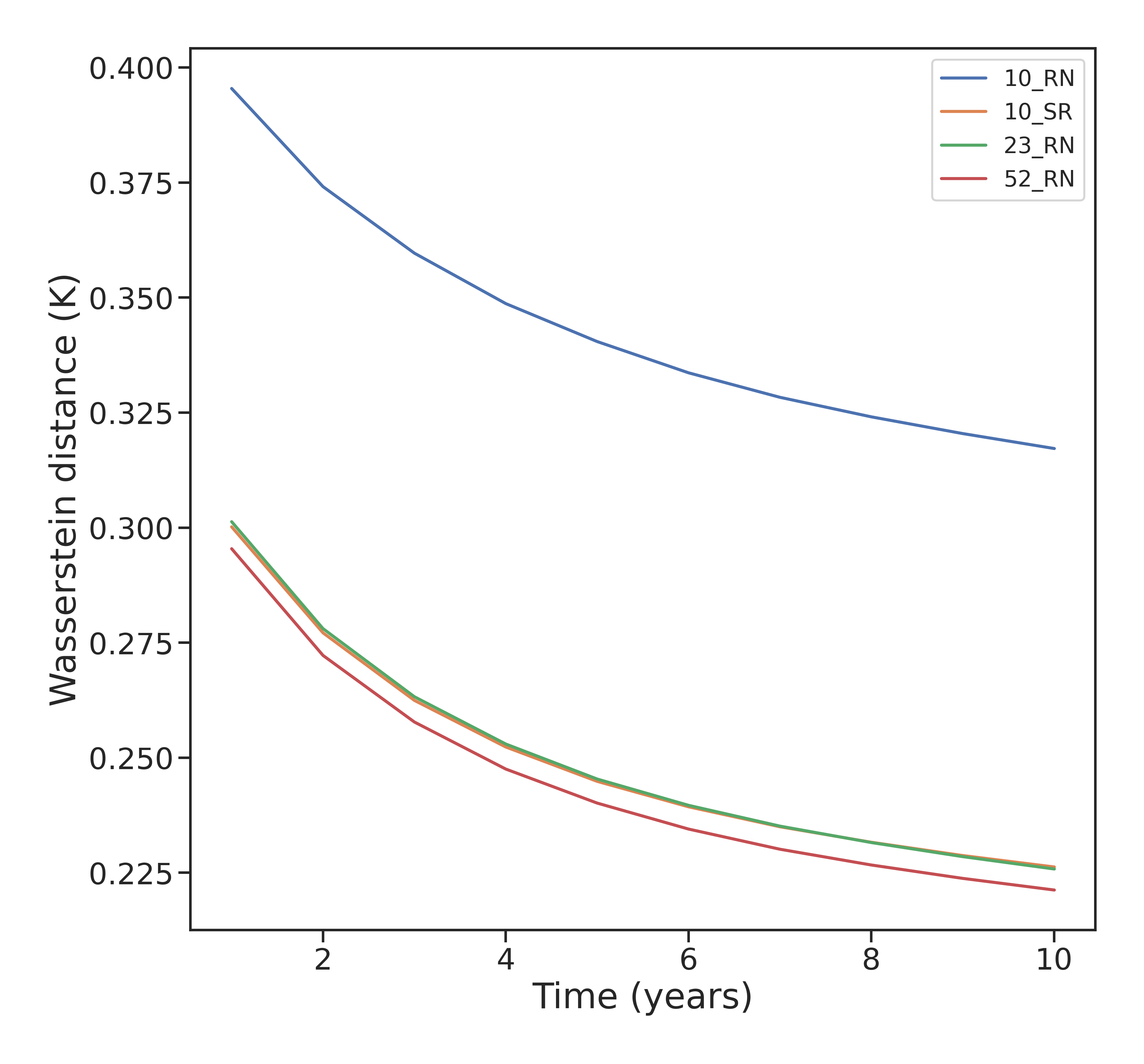}}
	\subfloat[\label{fig:wasserstein_precipitation}]{\includegraphics[width=0.48\textwidth]{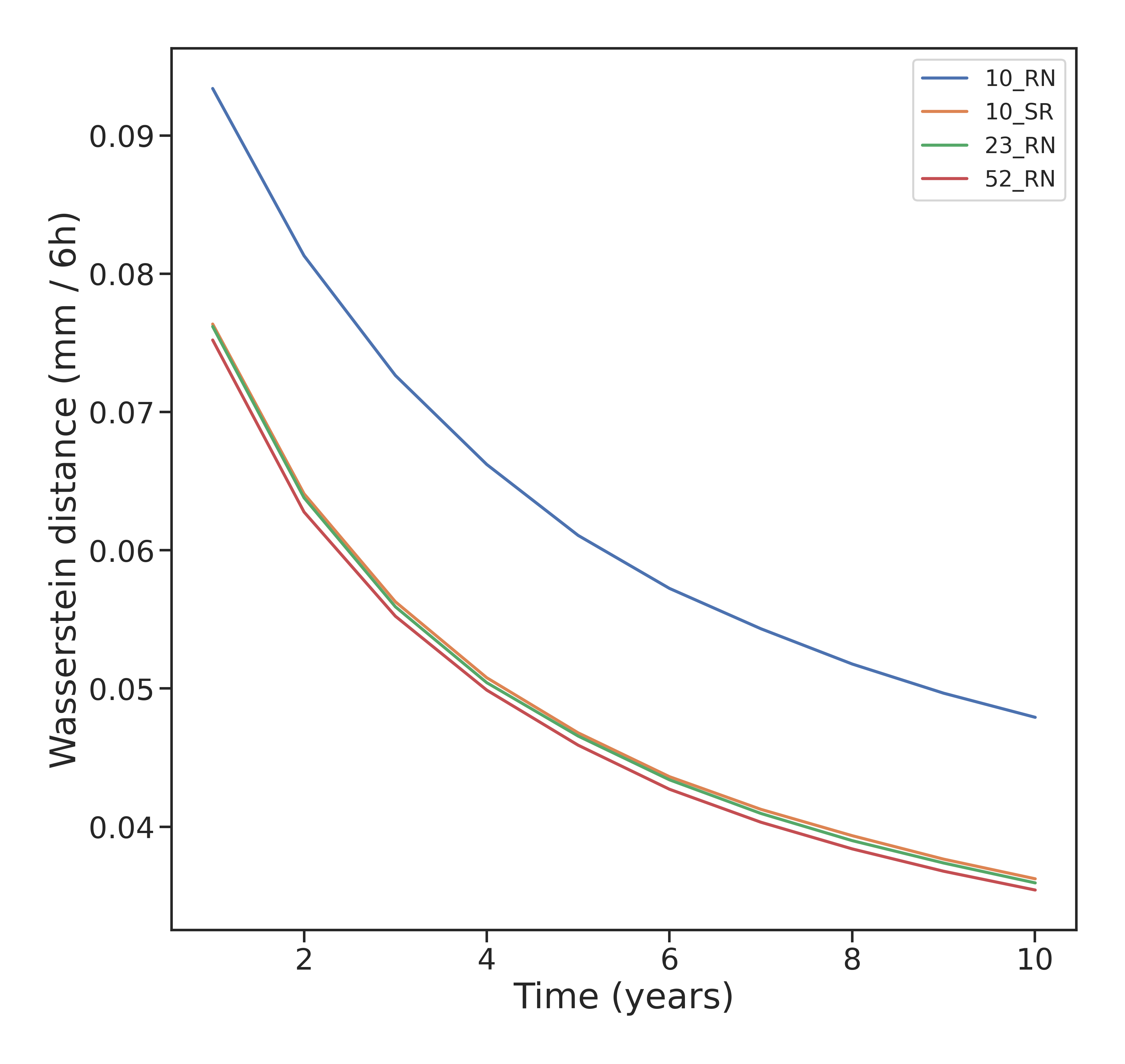}}
	\caption{Area-weighted mean grid point Wasserstein distances for (a) temperature and (b) precipitation for the competitor ensembles $e^{\text{competitor}}$ relative to the control ensemble $e^{\text{control}}$, for constant SST fields over 10 years, after 100 years of changing climate integration. Errors at reduced precision relative to double precision are generally small, with stochastic rounding dramatically improving the performance. The WD can be considered as an upper limit on the difference in the expected value between two distributions by Eq. \ref{eq:WDdefn}. } 
	\label{fig:wasserstein}
\end{figure*}

\noindent The mean Wasserstein distances (area-weighted average over all grid points) can be seen in Fig. \ref{fig:wasserstein} for both the global surface temperature and the total precipitation. For both variables, the distribution results quantified by the Wasserstein distance mirror the transient time-series effects that we demonstrated in the previous section (e.g. Fig. \ref{fig:single_timeseries} ) Firstly, the single precision solution reproduces the double precision solution with high fidelity. The Wasserstein distance error (i.e. the difference between the double solution - red line in Fig \ref{fig:wasserstein} - and the reduced precision solution), after 10 years is $4.6 \times 10^{-3}$ K for temperature and $5.2 \times 10^{-4}$ mm/6h for precipitation. Secondly, the half precision solution with round to nearest - despite being our worse performing solution - exhibits generally reasonable performance, well-approximating the double precision solution with WD error of 0.1 K and 0.012 mm/6h for temperature and precipitation respectively. Again the addition of stochastic rounding to the half precision solution vastly improves the performance; for temperature the WD error is $5 \times 10^{-3}$ K and for precipitation, $8 \times 10^{-4}$ mm/6h, highly comparable to the single precision behaviour. To emphasize the interpretation of the Wasserstein distance which may be unfamiliar to readers, this means that after 110 years of numerical integration (100 with time-variable SSTs, 10 with time-constant) the difference in (e.g.) the expected value of the climate global mean surface temperature between the double precision solution and the half precision solution with stochastic rounding is $\leq 5 \times 10^{-3}$ K. Referring back to the two original questions that we posed at the start of this work, we can see that the climate change signal can be well reproduced at reduced precision (highly accurate at single precision, small errors at half precision) and that the errors that we see at half precision can be significantly reduced by the addition of stochastic rounding (half precision with stochastic rounding is comparable to double precision).

\section{Impact of Stochastic Rounding and Mixed precision }\label{sec4_SRMP}

It is evident from the preceding analysis that the climate change signal can be well-reproduced at reduced precision, with marked improvements if we use stochastic rounding, rather than a round-to-nearest approach. The question then becomes \textit{why} stochastic rounding improves the low precision solution? Can we identify parts of the model that fail when we go to 10 significand bits with round to nearest, but are fixed by the addition of stochastic rounding? \newline
\begin{figure*}
	\centering
	\begin{subfigure}[b]{0.475\textwidth}
		\centering
		\includegraphics[width=\textwidth]{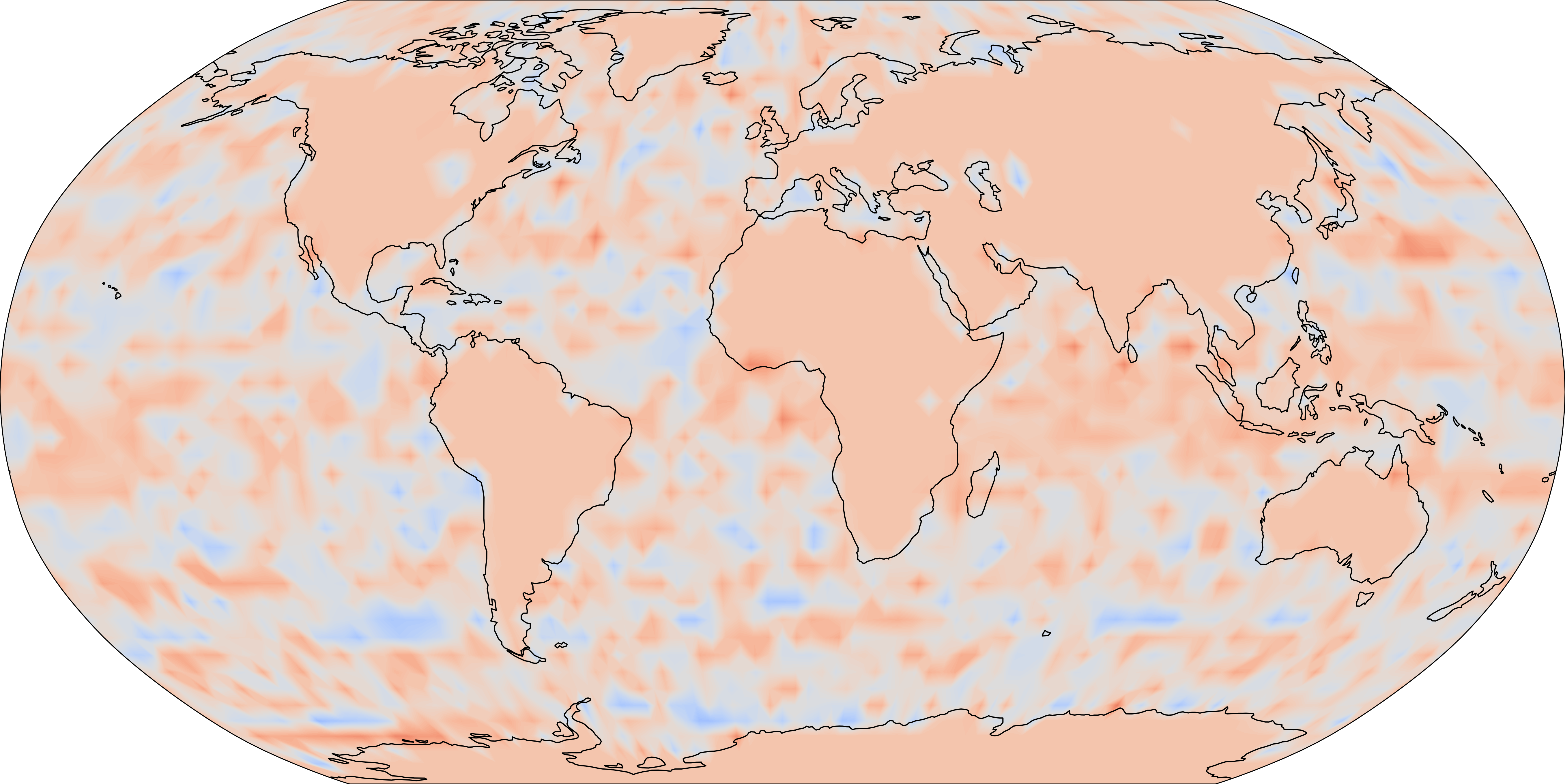}
		\caption*{Sea, RN}    
		\label{fig:mean and std of net14}
	\end{subfigure}
	\begin{subfigure}[b]{0.475\textwidth}  
		\centering 
		\includegraphics[width=\textwidth]{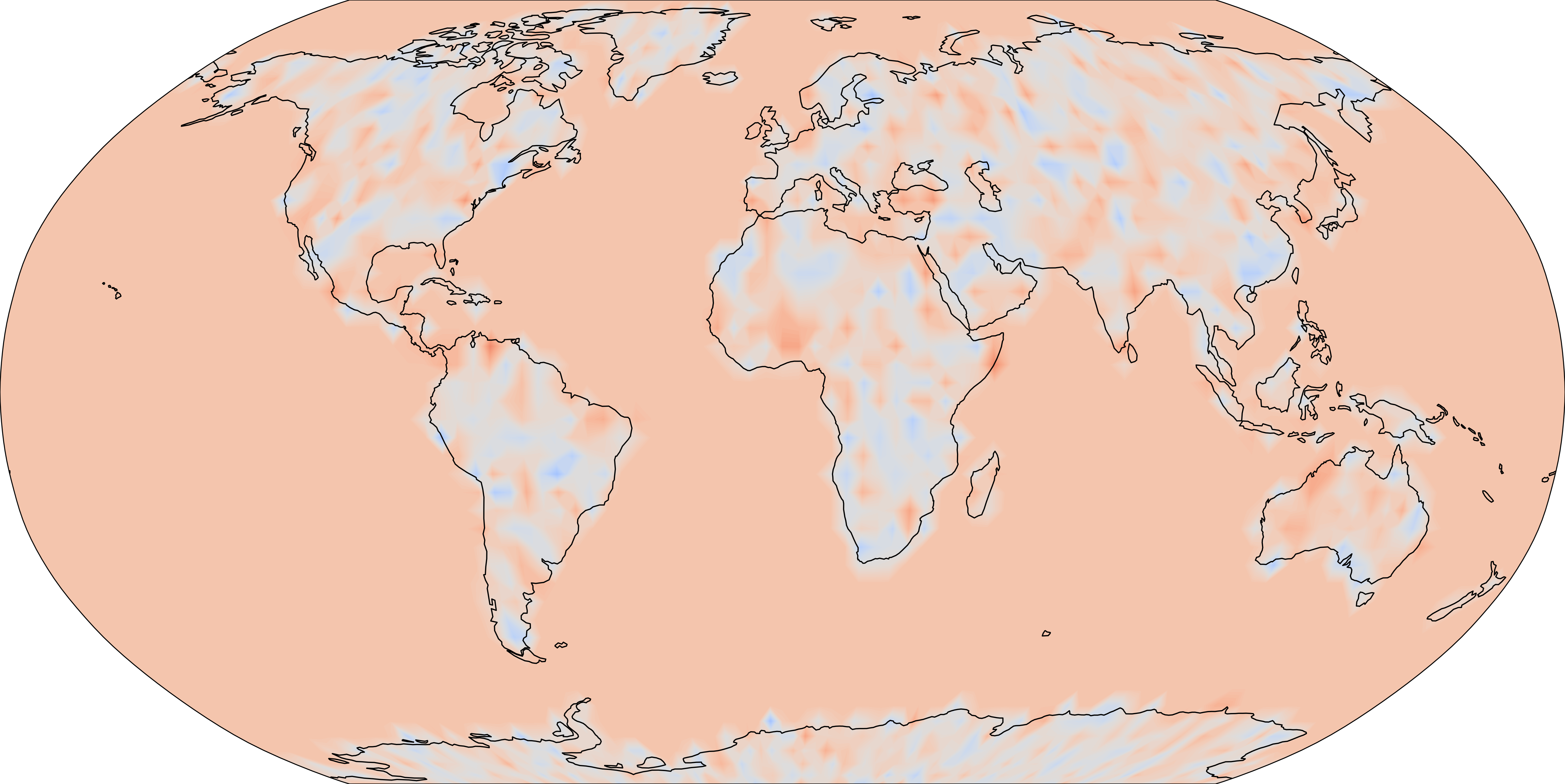}
		\caption*{Land, RN}  
		\label{fig:mean and std of net24}
	\end{subfigure}
	\begin{subfigure}[b]{0.475\textwidth}   
		\centering 
		\includegraphics[width=\textwidth]{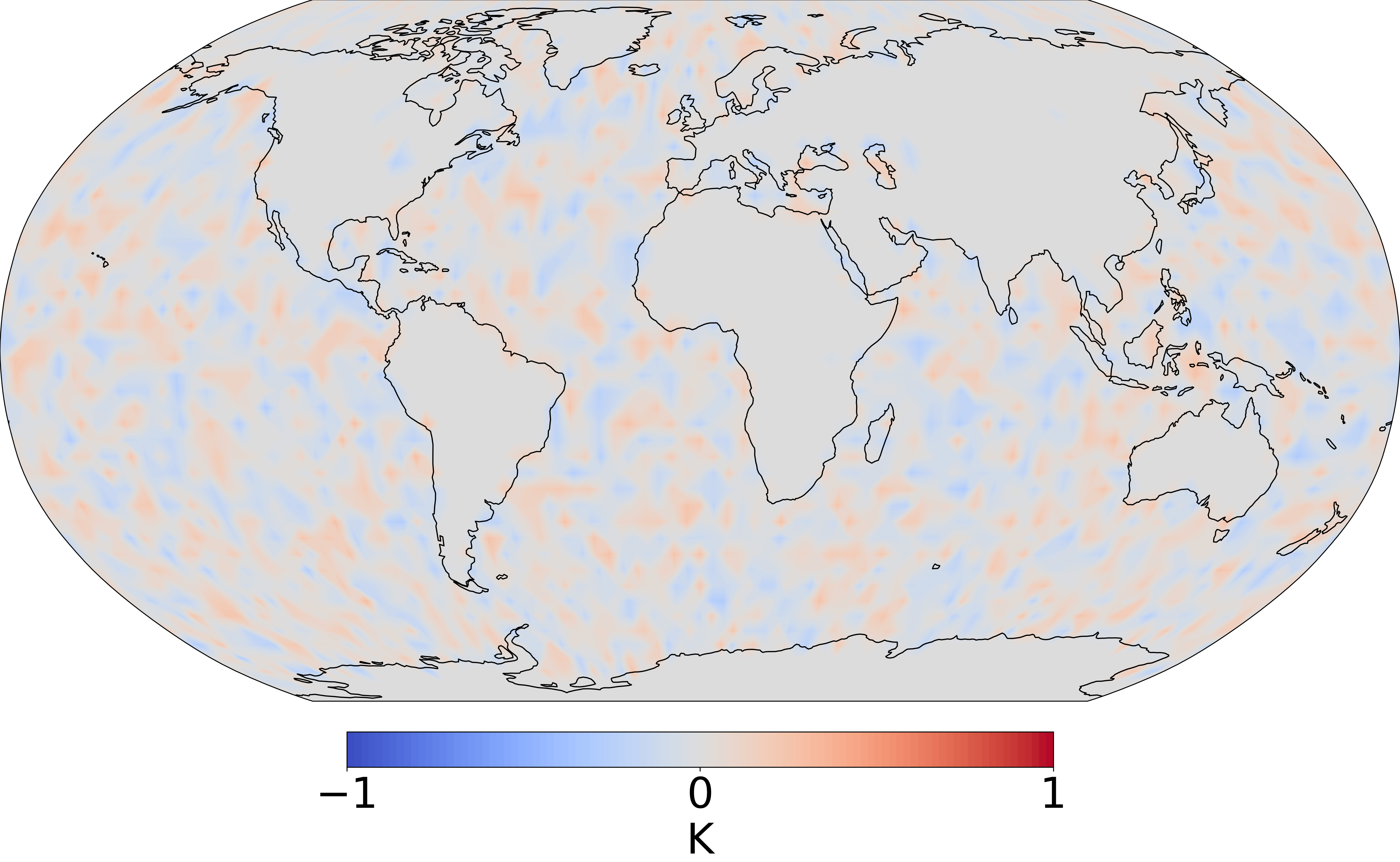}
		\caption*{Sea, SR}  
		\label{fig:mean and std of net34}
	\end{subfigure}
	\begin{subfigure}[b]{0.475\textwidth}   
		\centering 
		\includegraphics[width=\textwidth]{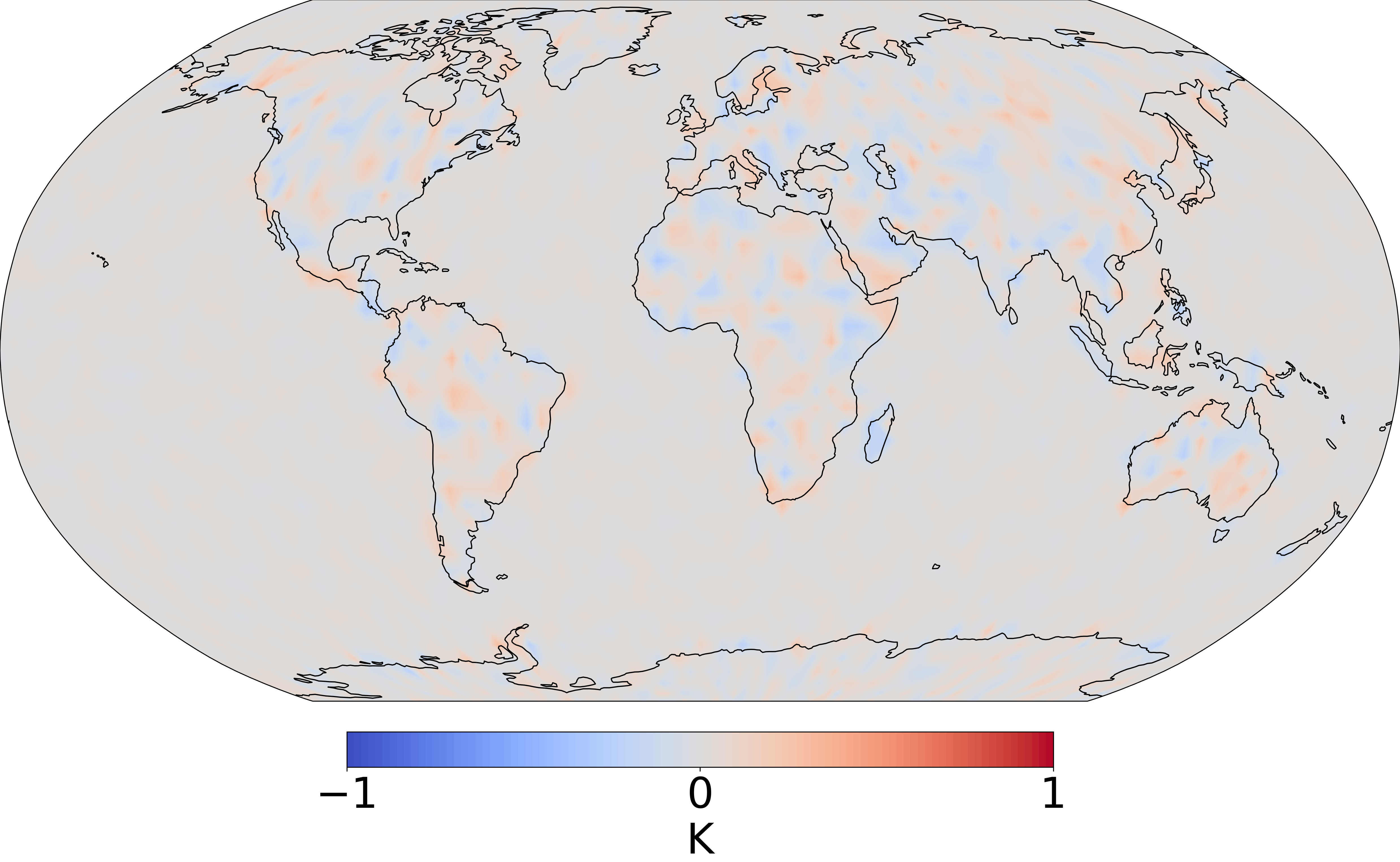}
		\caption*{Land, SR} 
		\label{fig:mean and std of net44}
	\end{subfigure}
	\caption{Mean error - relative to the double precision solution - in the interpolated sea (left column) and land (right column) surface temperatures, averaged over a 10 year integration period, at half precision using round-to-nearest (top row) and stochastic rounding (bottom row).  For the interpolated sea surface temperatures, round-to-nearest displays a strong, uniform, biased error in the temperature field over the land, along with substantial, variable errors over the sea. In contrast, stochastic rounding enjoys accurate interpolation over both land and sea. The corollary is true for the interpolated land surface temperatures. } 
	\label{fig:mean and std of nets}
\end{figure*}

\noindent In general it is difficult to identify all precision bottlenecks in a model, especially for those models which are larger and more complex. An example bottleneck that we have pinpointed in the SPEEDY model where the low precision RN solution starts to incur larger errors which are mitigated by using stochastic rounding is in the interpolation of climatological sea-land fields, i.e. the interpolation of the monthly-mean surface temperatures for land and sea down to a daily cadence. After each day of integration, the SPEEDY model exchange fluxes between the atmosphere and the land/sea. This necessitates both a land surface model and a sea surface model. Each of these surface models involve taking an observed monthly mean surface field from file and then interpolating to obtain a daily surface field for the land/sea surface temperatures. These daily surface fields then serve as boundary conditions. The error in the interpolated daily temperature surface fields for both land and sea for our reduced precision solutions, relative to the double precision solution, averaged over an integration period of 10 years are presented in Fig. \ref{fig:mean and std of nets}, for both round-to-nearest and stochastic rounding. It is immediately evident that the interpolation using half precision, round-to-nearest induces a large systematic error for both the sea and land. Consider the interpolation of the sea surface temperature. We know that the values over land are zero (273 K) by design. We therefore require that the interpolated land values are also zero. However the half precision round-to-nearest solution induces a systematic error that strongly drives the land values away from zero. In contrast when using stochastic rounding the mean interpolated values over the land are accurate. The corollary is also true for interpolating the land surface temperature fields; we expect the sea values to be zero but round-to-nearest induces a strong bias which is cured by using stochastic rounding. We should emphasize that the improved performance of stochastic rounding is evident when taking an average over multiple days; for an individual interpolated day SR does exhibit errors over the zero-valued regions, but these errors are not a systematic bias as in the RN case, instead being random and unbiased. The statistical properties of SR ensure that when averaged over long timescales these errors are mean zero. In addition to the accurate behaviour over the zero valued regions, SR exhibits a generally improved performance in the regions where the surface field is non-zero e.g. over the sea for the sea surface temperature.   \newline

\begin{figure}
 	\includegraphics[width=\columnwidth]{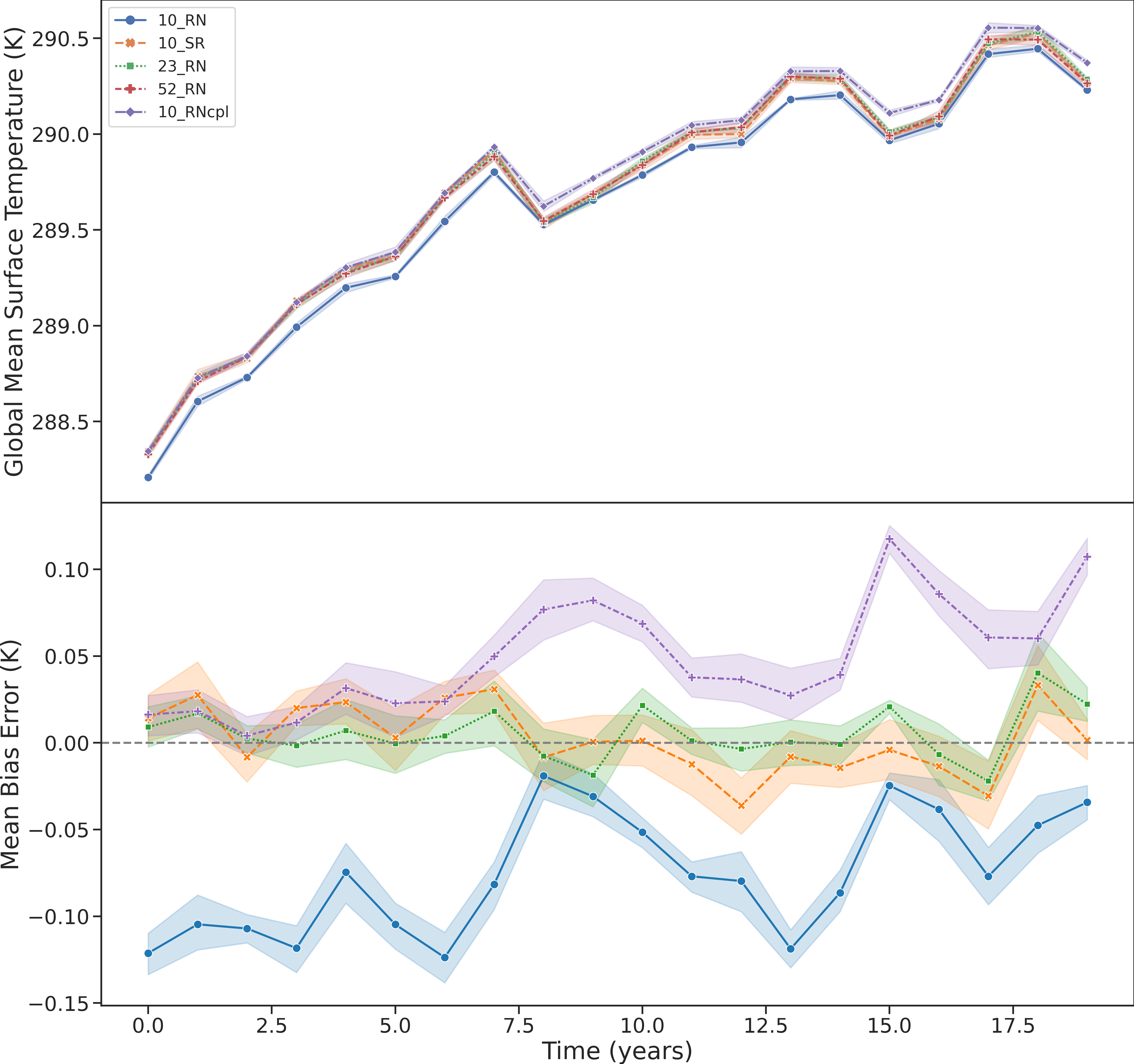}
 	\caption{As Fig. \ref{fig:ensemble_timeseries}, but over a 20 year integration period when the climate is changing most rapidly, and with the coupling mixed precision solution also plotted (10$\_$RNcpl). The mixed precision solution shows superior behaviour to the half precision solution at early times, but diverges from the ground truth at later times.}
 	\label{fig:mixed_prec_cpl}
 \end{figure}

 \noindent We can explore the impact of the interpolation error at low precision on the resultant half precision surface temperature/precipitation solutions by constructing an additional mixed float precision solution where the overall model is run at half precision as before, but the interpolation of the sea/land fields is done at double precision. All the remaining components of the model e.g. initialization, physics and dynamics tendencies, Fourier transforms, spectral/grid point transforms etc. use 10 significand bits. In this way we can think of these mixed precisions as the half precision solutions which have been selectively augmented. The results for the global surface temperature are presented in Fig \ref{fig:mixed_prec_cpl}, which is an analogous to the ensemble time series of  Fig. \ref{fig:ensemble_timeseries}, now with the additional mixed precision solution included (10$\_$RNcpl), and only showing the first 20 years of integration.
We can see that the mixed precision solution improves the accuracy at early times. Over the initial 10 year period where the temperature gradients are largest and the climate is changing most rapidly, the MBE for the mixed precision solution is $3.3 \times 10^{-2}$ K, compared to the 10$\_$RN with MBE $ \sim 8.9 \times 10^{-2}$ K. However at later times the solution does start to diverge and is generally outperformed by the SR solution, suggesting that there are other sections of the model which are also improved by stochastic rounding. Similarly, the mixed precision solution has only a modest effect on the precipitation time series, suggesting that other biased calculations are responsible for these errors. \newline

\section{Discussion}\label{sec5_discus}

Since stochastic rounding fundamentally trades short term accuracy for long term statistics, it seems reasonable to suggest that SR may favourably lend itself to long timescale climate simulations at reduced precision. This is indeed what we have observed; whilst a naive reduction from double precision to half precision using deterministic rounding results in a solution which is generally accurate with some small errors, the addition of stochastic rounding significantly improves the performance to be almost on parity with the full Float64 solution. This definitively answers the two questions we posed in the introduction at the start of this work. Whilst it is difficult to extrapolate from the simple atmospheric model we have used to more complex state-of-the-art and next generational models, the strong performance at reduced precision suggests that the use of reduced precision in more complex simulations could be a fruitful avenue to explore, at least in some parts of the model. At higher resolutions it could be the case that the derivatives are too small to be naively represented at reduced precision. In this case it would be necessary to either scale the values into an appropriate range or else  perform the computations in high precision (e.g. Float32), and then stochastically round the numbers to 16 bits before they are transported to a different part of the computer. Indeed, most computing energy in a modern supercomputer is spent transporting data from one node to another or to memory \cite{exacompute}. By using stochastic rounding for transport rather than explicitly calculating dervatives we might be able to get the advantages of reduced precision in terms of savings of machine energy, but without having to degrade the calculation of derivatives. SR is also particularly advantageous for application on these large computers since any incurred computational cost is likely to be dwarfed by the expense of transporting data and so the cost of 10\_RN and 10\_SR would be highly comparable. \newline

\noindent We have also presented a demonstrative example area of the SPEEDY model where half precision fails but stochastic rounding succeeds - the interpolation of the sea/land fields. For simpler and smaller codes it is possible to identify such precision bottlenecks. In this case once could rescale the relevant equations to ensure that the float operations can be described by the necessary number format c.f. dynamic range and precision (see e.g. \cite{kloewer2022a}). Conversely, for larger, more complex codes rescaling equations and refactoring code can be prohibitively difficult. Instead, as evidenced in this work, it could be more straightforward to instead use stochastic rounding as a drop-in replacement for deterministic rounding at reduced float precision. Indeed, the Graphcore Colossus Mk2 IPU chips \cite{graphcore} which implement stochastic rounding at the hardware level, allow for just such an approach where the user can simply swap out deterministic for stochastic rounding. If other chip-makers follow the same example it will become much more feasible to implement reduced precision stochastic rounding on large-scale models. More generally, inefficiencies or small bugs that might be overlooked or gotten away with when using double precision can become important at reduced precision. For instance, it would be straightforward to introduce a patch to SPEEDY that enforces the mean interpolated sea/ice fields over land to be zero, irrespective of what the calculated interpolated value is. An apparent failure of the model when going to reduced precision is then not necessarily due to some fundamental loss of information, but can instead be due to the underlying code to being structured in a way that is sub-optimal for reduced precision methods. Whilst this does introduced an additional `offline' overhead -  i.e. one cannot simply change the precision, but may need top optimize the code for that precision - this could be heavily offset by the computational gains that could be made through such optimizations. \newline

\noindent The development of more modern implementations of global atmospheric models such as SPEEDY will also facilitate the exploration and prototyping of 16-bit climate models. For example, SpeedyWeather.jl \cite{SpeedyJL} - which is currently under development - is a climate model analogous to SPEEDY, but written in a way that is fully type-flexible to enable arbitrary number formats for performance and analysis simultaneously, with support for hardware accelerated low precision. This means the model development is precision-agnostic, which allows the common problems of dynamic range and critical precision loss  - often incurred from using low-precision number formats - to be directly addressed. Such a model will also be fully differentiable via automatic differentiation, with entire parts of the model able to be replaced with artificial neural networks. This hybrid structure enables an improved representation of climactic processes (by e.g. training against higher resolution simulations) and the fitting of models to observational data. The ability to deploy such a model directly on hardware with inbuilt stochastic rounding support would be a highly informative with respect to the construction of large scale 16-bit climate models.

\section{Conclusions}
We have demonstrated that we can model a changing climate system using reduced precision floats, subject to minor errors with respect to the double precision solution. Whilst the global atmospheric model used in this work is fundamentally an approximation to higher-order state-of-the-art climate models, the effectiveness of reduced precision techniques motivates the development and further exploration of reduced precision in more complex simulations, at least in some parts of the model. The addition of stochastic rounding significantly improves the half precision performance to be comparable to the single precision solution, with only very minor differences to the full double precision solution. This suggests that the information content of the higher order bits is low and that significant performance gains could be realised through selective use of reduced precision and stochastic rounding.

\section*{Code}
This implementation of SPEEDY is based off original work by Saffin and Hatfield \cite{SPEEDY}, who developed a Fortran model based on work by Kucharski, Molenti et al. \cite{Kucharski}, modified to include a reduced precision emulator \cite{rpeV5}. This model was in turn amended by Pax21 to include stochastic rounding in the emulator \url{https://github.com/eapax/speedy}.
The exact SPEEDY model used in this work is based of a branch from Pax21 to which some small refactoring modifications have been made.

\section*{Acknowledgments}
This project has received funding from the European Research Council (ERC) under the European Union’s Horizon 2020 research and innovation programme (Grant No 741112).

\bibliographystyle{ieeetr}
\bibliography{refs}

\begin{thebibliography}{10}

\bibitem{Chantry2019}
M.~{Chantry}, T.~{Thornes}, T.~{Palmer}, and P.~{D{\"u}ben}, ``{Scale-Selective
  Precision for Weather and Climate Forecasting},'' {\em Monthly Weather
  Review}, vol.~147, pp.~645--655, Feb. 2019.

\bibitem{Hatfield2019}
S.~Hatfield, M.~Chantry, P.~D\"{u}ben, and T.~Palmer, ``Accelerating
  high-resolution weather models with deep-learning hardware,'' in {\em
  Proceedings of the Platform for Advanced Scientific Computing Conference},
  PASC '19, (New York, NY, USA), Association for Computing Machinery, 2019.

\bibitem{Lang2021}
S.~T.~K. Lang, A.~Dawson, M.~Diamantakis, P.~Dueben, S.~Hatfield,
  M.~Leutbecher, T.~Palmer, F.~Prates, C.~D. Roberts, I.~Sandu, and N.~Wedi,
  ``More accuracy with less precision,'' {\em Quarterly Journal of the Royal
  Meteorological Society}, vol.~147, no.~741, pp.~4358--4370, 2021.

\bibitem{ECMWF}
C.~Maass.
  \url{https://confluence.ecmwf.int/display/FCST/Implementation+of+IFS+Cycle+47r2}.
\newblock Accessed: 2021-11-04.

\bibitem{Vana2016}
F.~Vana, P.~Düben, S.~Lang, T.~Palmer, M.~Leutbecher, D.~Salmond, and
  G.~Carver, ``Single precision in weather forecasting models: An evaluation
  with the ifs,'' {\em Monthly Weather Review}, vol.~145, 12 2016.

\bibitem{MetOffice}
R.~Gilham, ``32-bit physics in the unified model.''
  \url{https://digital.nmla.metoffice.gov.uk/download/file/IO_951e52e5-6698-485e-ad33-54d0a2b0ce99},
  2018.

\bibitem{SWISS}
S.~{Rudisuhli}, A.~{Walser}, and O.~{Fuhrer}, ``Cosmo in single precision,''
  2014.

\bibitem{Hopkins2020}
M.~{Hopkins}, M.~{Mikaitis}, D.~R. {Lester}, and S.~{Furber}, ``{Stochastic
  rounding and reduced-precision fixed-point arithmetic for solving neural
  ordinary differential equations},'' {\em Philosophical Transactions of the
  Royal Society of London Series A}, vol.~378, p.~20190052, Mar. 2020.

\bibitem{Rehm2021}
F.~{Rehm}, S.~{Vallecorsa}, V.~{Saletore}, H.~{Pabst}, A.~{Chaibi},
  V.~{Codreanu}, K.~{Borras}, and D.~{Kr{\"u}cker}, ``{Reduced Precision
  Strategies for Deep Learning: A High Energy Physics Generative Adversarial
  Network Use Case},'' {\em arXiv e-prints}, p.~arXiv:2103.10142, Mar. 2021.

\bibitem{Gupta2021}
R.~R. Gupta and V.~Ranga, ``Comparative study of different reduced precision
  techniques in deep neural network,'' in {\em Proceedings of International
  Conference on Big Data, Machine Learning and their Applications} (S.~Tiwari,
  E.~Suryani, A.~K. Ng, K.~K. Mishra, and N.~Singh, eds.), (Singapore),
  pp.~123--136, Springer Singapore, 2021.

\bibitem{Graphcore2022}
B.~{Noune}, P.~{Jones}, D.~{Justus}, D.~{Masters}, and C.~{Luschi}, ``{8-bit
  Numerical Formats for Deep Neural Networks},'' {\em arXiv e-prints},
  p.~arXiv:2206.02915, June 2022.

\bibitem{NVIDIA}
``{Training With Mixed Precision}.''
  \url{https://docs.nvidia.com/deeplearning/performance/mixed-precision-training/index.html},
  06 2021.
\newblock Accessed: 2021-11-04.

\bibitem{TF}
``Accelerating ai performance on 3rd gen intel xeon scalable processors with
  tensorflow and bfloat16.''
  \url{https://blog.tensorflow.org/2020/06/accelerating-ai-performance-on-3rd-gen-processors-with-tensorflow-bfloat16.html},
  06 2020.
\newblock Accessed: 2021-11-04.

\bibitem{Paxton2021}
E.~A. Paxton, M.~Chantry, M.~Klöwer, L.~Saffin, and T.~Palmer, ``Climate
  modeling in low precision: Effects of both deterministic and stochastic
  rounding,'' {\em Journal of Climate}, vol.~35, no.~4, pp.~1215 -- 1229, 2022.

\bibitem{IEEEstandard}
``{IEEE Standard for Floating-Point Arithmetic},'' {\em IEEE Std 754-2019
  (Revision of IEEE 754-2008)}, pp.~1--84, 2019.

\bibitem{Higham2002}
N.~J. Higham, {\em Accuracy and Stability of Numerical Algorithms}.
\newblock USA: Society for Industrial and Applied Mathematics, 2nd~ed., 2002.

\bibitem{Connolly2021}
M.~P. Connolly, N.~J. Higham, and T.~Mary, ``Stochastic rounding and its
  probabilistic backward error analysis,'' {\em SIAM Journal on Scientific
  Computing}, vol.~43, no.~1, pp.~A566--A585, 2021.

\bibitem{SR1}
G.~E. Forsythe, ``Reprint of a note on rounding-off errors,'' {\em SIAM
  Review}, vol.~1, no.~1, pp.~66--67, 1959.

\bibitem{SR2}
T.~E. Hull and J.~R. Swenson, ``Tests of probabilistic models for propagation
  of roundoff errors,'' {\em Commun. ACM}, vol.~9, p.~108–113, feb 1966.

\bibitem{Gupta2015}
S.~{Gupta}, A.~{Agrawal}, K.~{Gopalakrishnan}, and P.~{Narayanan}, ``{Deep
  Learning with Limited Numerical Precision},'' {\em arXiv e-prints},
  p.~arXiv:1502.02551, Feb. 2015.

\bibitem{Na2017}
T.~Na, J.~H. Ko, J.~Kung, and S.~Mukhopadhyay, ``On-chip training of recurrent
  neural networks with limited numerical precision,'' in {\em 2017
  International Joint Conference on Neural Networks (IJCNN)}, pp.~3716--3723,
  2017.

\bibitem{Wang2018}
N.~{Wang}, J.~{Choi}, D.~{Brand}, C.-Y. {Chen}, and K.~{Gopalakrishnan},
  ``{Training Deep Neural Networks with 8-bit Floating Point Numbers},'' {\em
  arXiv e-prints}, p.~arXiv:1812.08011, Dec. 2018.

\bibitem{Xia2021}
L.~{Xia}, M.~{Anthonissen}, M.~{Hochstenbach}, and B.~{Koren}, ``{A Simple and
  Efficient Stochastic Rounding Method for Training Neural Networks in Low
  Precision},'' {\em arXiv e-prints}, p.~arXiv:2103.13445, Mar. 2021.

\bibitem{Fasi2021}
M.~Fasi and M.~Mikaitis, ``Algorithms for stochastically rounded elementary
  arithmetic operations in ieee 754 floating-point arithmetic,'' {\em IEEE
  Transactions on Emerging Topics in Computing}, vol.~9, no.~3, pp.~1451--1466,
  2021.

\bibitem{Croci2022}
M.~Croci and M.~B. Giles, ``{Effects of round-to-nearest and stochastic
  rounding in the numerical solution of the heat equation in low precision},''
  {\em IMA Journal of Numerical Analysis}, 04 2022.
\newblock drac012.

\bibitem{Jia2019}
Z.~{Jia}, B.~{Tillman}, M.~{Maggioni}, and D.~P. {Scarpazza}, ``{Dissecting the
  Graphcore IPU Architecture via Microbenchmarking},'' {\em arXiv e-prints},
  p.~arXiv:1912.03413, Dec. 2019.

\bibitem{graphcore}
``{IPU Hardware Overview}.''
  \url{https://docs.graphcore.ai/projects/ipu-programmers-guide/en/latest/about_ipu.html}.
\newblock Accessed: 2022-04-01.

\bibitem{CrociSRReview}
M.~Croci, M.~Fasi, N.~Higham, T.~Mary, and M.~Mikaitis, ``Stochastic rounding:
  implementation, error analysis and applications,'' {\em Royal Society Open
  Science}, vol.~9, 03 2022.

\bibitem{SPEEDYOG}
F.~Molteni, ``Atmospheric simulations using a gcm with simplified physical
  parametrizations. i: Model climatology and variability in multi-decadal
  experiments,'' {\em Climate Dynamics}, vol.~20, pp.~175--191, 01 2003.

\bibitem{SPEEDY}
L.~Saffin, S.~Hatfield, P.~D{\"u}ben, and T.~Palmer, ``Reduced precision
  parametrization: lessons from an intermediate complexity atmospheric model,''
  {\em Wiley}, February 2020.

\bibitem{rpeV5}
A.~{Dawson} and P.~D. {D{\"u}ben}, ``{rpe v5: an emulator for reduced
  floating-point precision in large numerical simulations},'' {\em
  Geoscientific Model Development}, vol.~10, pp.~2221--2230, June 2017.

\bibitem{RAW}
P.~Williams, ``A proposed modification to the robert–asselin time filter,''
  {\em Monthly Weather Review}, vol.~137, 08 2009.

\bibitem{Eyring2016}
V.~Eyring, S.~Bony, G.~A. Meehl, C.~A. Senior, B.~Stevens, R.~J. Stouffer, and
  K.~E. Taylor, ``Overview of the coupled model intercomparison project phase 6
  (cmip6) experimental design and organization,'' {\em Geoscientific Model
  Development}, vol.~9, no.~5, pp.~1937--1958, 2016.

\bibitem{ECEarth}
K.~{Wyser}, T.~{van Noije}, S.~{Yang}, J.~{von Hardenberg}, D.~{O'Donnell}, and
  R.~{D{\"o}scher}, ``{On the increased climate sensitivity in the EC-Earth
  model from CMIP5 to CMIP6},'' {\em Geoscientific Model Development}, vol.~13,
  pp.~3465--3474, Aug. 2020.

\bibitem{Dudley}
R.~M. Dudley, ``The speed of mean glivenko-cantelli convergence,'' {\em The
  Annals of Mathematical Statistics}, vol.~40, no.~1, pp.~40--50, 1969.

\bibitem{Vissio2020}
G.~{Vissio}, V.~{Lembo}, V.~{Lucarini}, and M.~{Ghil}, ``{Ranking IPCC Models
  Using the Wasserstein Distance},'' {\em arXiv e-prints}, p.~arXiv:2006.09304,
  June 2020.

\bibitem{exacompute}
P.~Kogge and J.~Shalf, ``Exascale computing trends: Adjusting to the "new
  normal"' for computer architecture,'' {\em Computing in Science and
  Engineering}, vol.~15, no.~6, pp.~16--26, 2013.

\bibitem{kloewer2022a}
M.~Klöwer, S.~Hatfield, M.~Croci, P.~Düben, and T.~Palmer, ``Fluid
  simulations accelerated with 16 bits: Approaching 4x speedup on a64fx by
  squeezing shallowwaters.jl into float16,'' {\em Journal of Advances in
  Modelling Earth Systems}, vol.~14, no.~2, 2022.

\bibitem{SpeedyJL}
``Speedyweather.jl.'' \url{https://github.com/milankl/SpeedyWeather.jl}, 2022.

\bibitem{Kucharski}
F.~Kucharski, F.~Molteni, M.~P. King, R.~Farneti, I.-S. Kang, and L.~Feudale,
  ``On the need of intermediate complexity general circulation models: A
  “speedy” example,'' {\em Bulletin of the American Meteorological
  Society}, vol.~94, no.~1, pp.~25 -- 30, 2013.

\end{thebibliography}

\end{document}